\documentstyle[12pt,prb,aps,epsf]{revtex}

\begin{document}
\draft
\title{Semiclassical Analysis of Extended Dynamical Mean Field Equations}
\author{Sergey Pankov, Gabriel Kotliar and Yukitoshi Motome$^\dagger$}
\address{Center for Materials Theory, Department of Physics and Astronomy,\\
Rutgers University, Piscataway, New Jersey 08854}
\date{\today}
\maketitle

\begin{abstract}
The extended Dynamical Mean Field Equations (EDMFT) are analyzed using
semiclassical methods for a model describing an interacting fermi-bose
system. We compare the semiclassical approach with the exact QMC (Quantum
Montecarlo) method. %
We found the transition to an ordered state to be of the first order for any
dimension below four.
\end{abstract}

\section{Introduction}

The dynamical mean field theory (DMFT) , a recently developed many body
approach to strongly correlated electron systems has been very successful in
unraveling non perturbative problems such as the Mott metal to insulator
transition.\cite{Georges1996} In spite of its many successes, this
technique has several limitations resulting from its single site character
and from the lack of feedback of the non local collective excitations on the
one particle spectra. Several approaches are being pursued to extend the
scope of the DMFT method. In this paper we explore  
an extension of the  DMFT
method (EDMFT) \cite{sachdev,Kajueter96,Si96} which maintains a local self
energy while incorporating  feedback effects of the charge and spin dynamics
in the one electron properties.  
This method gives rise to quantum
impurity problems with fermionic and bosonic baths that need to be solved
self consistently.
This  method has already   been applied to wide class of models, such
as the spin fermion model, \cite{Si0012373}
fermions interacting with
long range (Coulomb) electron-electron interaction,\cite{Chitra9903180}
electron-phonon systems\cite{Motome2000} and frustrated magnets.\cite{georges}

The EDMFT equations are more involved than the conventional DMFT
equations because they involve a solution of a self consistency problem in
an additional bosonic sector and only recently a full numerical analysis of
the self consistency conditions of EDMFT was carried out.\cite{Motome2000}
The interpretation of the EDMFT instabilities is also  not
as straightforward as in DMFT because bosonic and fermionic propagators
involve very different regions of momentum space, and a formulation
of EDMFT for ordered phases was only obtained recently.\cite{chitra2}  

The purpose of this paper is to develop the EDMFT approach further by
analyzing several aspects of this method: a)  
We implement  a semiclassical technique for its solution
\cite{semiclassical,Freericks1993,Millis1996} 
and compare its results to the earlier QMC
study\cite{Motome2000} to test its accuracy.
We show that the analytic treatment is in satisfactory agreement
with exact (QMC) results in the high temperature regime of 
the three dimensional (3d) model and provides
analytic expressions for various physical quantities. 
b) We extend
this  study to the case of two dimensional (2d)
phonons, which had not been treated in ref.\cite{Motome2000}
We demonstrate that  in the 2d case
the EDMFT  treatment  
at finite temperatures, if it produces an ordering
transition it  is necessarily 
of the first order.
This analysis applies to a very general class of models including
those used in ref. \cite{Si0012373} 

We also analyze the EDMFT equations in the ordered phase, 
\cite{chitra2} for a simple spin model.  
This analysis   clarifies 
the strengths and the limitations of the
EDMFT approach, in a very simple setting.

The paper is organized as follows. In section \ref{sec_spe} we write the
Fermion Boson model and the extended DMFT equations. Then we describe the
semiclassical strategy for their analysis in both weak and strong
electron-phonon coupling. In section \ref{sec_3d} we present results of
solving the saddle point equations for 3d phonons coupled to electrons in
different regimes and discuss the agreement with results in QMC approach. In
section \ref{sec_2d} we describe the results for 2d phonons. If the
electrons are fully integrated out the semiclassical treatment of EDMFT has
to reduce to a mean field theory in classical statistical mechanics. In
section \ref{tc} we compare EDMFT with other classical mean field
treatments such as the Weiss mean field approach and the Bloch Langer
method.\cite{BlochLanger}  The   stability analysis of the EDMFT   
theory is carried out in Appendix \ref{stability}.

\section{Model and semiclassical approximation}

\label{sec_spe}

The model under consideration is described by the lattice Hamiltonian:

\begin{equation}  \label{ham}
H=H_{el}+H_{ph}+H_{el-ph}
\end{equation}
where

\begin{equation}  \label{h_el}
H_{el}= -\sum\limits_{ij,\sigma} t_{ij}c_{i\sigma}^{\dagger}c_{j\sigma}
\end{equation}

\begin{equation}  \label{h_ph}
H_{ph}= \sum\limits_i{\frac{p_i^2}{2M}}- \sum\limits_{ij}{\frac{J_{ij}}{2}}
x_ix_j
\end{equation}

\begin{equation}  \label{h_el-ph}
H_{el-ph}= \sum\limits_{i\sigma} \lambda x_i
(c_{i\sigma}^{\dagger}c_{i\sigma}-{\frac{1}{2}})
\end{equation}

The first term describes free electrons, $c^{\dagger}_{i\sigma}$ 
($c_{i\sigma}$) creates (annihilates) an electron with spin $\sigma$ on
a site $i$. The second term describes nonlocal (dispersive) phonons,
$x_i$ and $p_i$ are canonical variables. The last term couples 
the fermionic and the bosonic degrees of freedom. We consider 
a half filled system of fermions. 

The second term could alternatively be written as:

\begin{equation}
H_{ph}= \sum\limits_q\omega_q(a_q^{\dagger}a_q+{1\over2})
\end{equation}
where $a_q$, $a^{\dagger}_q$ are related to the phonon field by
$x_q=(2 M\omega_q)^{-{1\over2}}(a_q+a^{\dagger}_{-q})$ and 
$\omega_q^2=J_q/M$. Dispersion (momentum dependence) of the boson 
frequency $\omega_q$ stems from nonlocal character of $J_{ij}$. 
The local limit of $J_{ij}$ corresponds to the Holstein 
model\cite{holstein59}, 
so the model under consideration is an
extension of the Holstein model to dispersive phonons. 

The extended DMFT equations for this model \cite{Motome2000} are a set of
equations for Weiss functions $G_{0\sigma }^{-1}(i\omega _{n})$ and $
D_{0}^{-1}(i\omega _{n})$:

\begin{equation}  \label{selfcons.G}
G_{\sigma}[G_0,D_0](i\omega_n)= \sum\limits_q\left[i\omega_n-t_q
+G_{\sigma}[G_0,D_0]^{-1}(i\omega_n)- G_{0\sigma}^{-1}(i\omega_n)\right]^{-1}
\end{equation}

\begin{equation}  \label{selfcons.D}
D[G_0,D_0](i\omega_n)= \sum\limits_q\left[M(i\omega_n)^2-J_q
+D[G_0,D_0]^{-1}(i\omega_n)-D_0^{-1}(i\omega_n)\right]^{-1}
\end{equation}
where full Green's functions $G_{\sigma}(i\omega_n)$ and $D(i\omega_n)$ are
expressed through $G_{0\sigma}^{-1}(i\omega_n)$ and $D_0^{-1}(i\omega_n)$ in
terms of the effective impurity action: 
\begin{equation}  \label{s_eff}
S_{eff}=\sum\limits_{\omega_n,\omega_m,\sigma}
c_{\sigma}^{\dagger}(i\omega_n+i\omega_m)
\left(G_{0\sigma}^{-1}(i\omega_n)\delta_{0,\omega_m} +\lambda
x(i\omega_m)\right)c_{\sigma}(i\omega_n) -{\frac{1}{2}}D_0^{-1}(i
\omega_m)x^2(i\omega_m)
\end{equation}

\begin{equation}  \label{defG}
G_{\sigma}(i\omega_n)= {\frac{\int{\cal D}[c_{\sigma}^{\dagger},c_{
\sigma},x] c_{\sigma}(i\omega_n)c_{\sigma}^{\dagger}(i\omega_n)
e^{-S_{eff}[c_{\sigma}^{\dagger},c_{\sigma},x]}}{\int{\cal D}
[c_{\sigma}^{\dagger},c_{\sigma},x]
e^{-S_{eff}[c_{\sigma}^{\dagger},c_{\sigma},x]}}}= \langle
c_{\sigma}(i\omega_n)c_{\sigma}^{\dagger}(i\omega_n) \rangle_{S_{eff}}
\end{equation}

\begin{equation}  \label{defD}
D_{\sigma}(i\omega_n)=- \langle x^2(i\omega_n)\rangle_{S_{eff}}
\end{equation}

Eqs. (\ref{selfcons.G}-\ref{defD}) in general have to be solved numerically.
In many cases though, a number of approximations reducing numerical work but
preserving a physical content of the problem is possible. One of the
approximations is in using a model density of states (DOS) 
for fermions and bosons, so that
momentum summations in EDMFT equations could be performed analytically. It
is convenient to chose semicircular electron DOS:

\begin{equation}
\rho _{el}(\epsilon )={\frac{2}{\pi W^{2}}}\sqrt{W^{2}-\epsilon ^{2}}
\label{el.dos}
\end{equation}
where $W$ is the electron band halfwidth. The particular choice of
semicircular electron DOS is qualitatively unimportant since we consider
half filled electron band. For phonons, on the contrary, the shape of the
phonon band near the bottom is crucial for temperatures smaller than the phonon
band width. For d-dimensional phonons the bottom of the band has 
$~\epsilon ^{\frac{d-2}{2}}$ singularity. 
That is why to represent 3d and 2d phonons we
chose semicircular and step-function-like phonon DOS respectively:

\begin{eqnarray}  \label{ph.3d.dos}
3d & \qquad\qquad & \rho_{ph}(\epsilon)= {\frac{2}{\pi\omega_1^2}}\sqrt{
\omega_1^2-(\epsilon-\omega_0)^2} \\ \label{ph.2d.dos}
2d & \qquad\qquad & \rho_{ph}(\epsilon)= {\frac{1}{2\omega_1}}
\theta\left(\omega_1^2-(\epsilon-\omega_0)^2\right)
\end{eqnarray}

After replacing summations over wave vector by integrations over energy, 
Eq. (\ref{selfcons.G}) and Eq. (\ref{selfcons.D}) read:

\begin{equation}  \label{enintG}
G_{\sigma}(i\omega_n)=\int d\epsilon {\frac{\rho_{el}(\epsilon)}{
\zeta-\epsilon}}
\end{equation}

\begin{equation}  \label{enintD}
D(i\omega_n)=\int d\epsilon {\frac{\rho_{ph}(\epsilon)}{\xi^2-\epsilon^2}}
\end{equation}
where $\zeta=
i\omega_n+G_{\sigma}^{-1}(i\omega_n)-G_{0\sigma}^{-1}(i\omega_n)$, $
\xi^2=M(i\omega_n)^2+D^{-1}(i\omega_n)-D_0^{-1}(i\omega_n)$; density of
states $\rho(\epsilon)\equiv{\frac{dq}{d\epsilon_q}}$. For electron $
\rho_{el}(\epsilon)$ and phonon $\rho_{ph}(\epsilon)$ DOS respectively $
\epsilon_q=t_q$ and $\epsilon_q^2=J_q$. For DOS defined in 
Eqs. (\ref{el.dos}-\ref{ph.2d.dos}) 
integrations over energy in Eqs. (\ref{enintG},\ref{enintD})
yield:

\begin{equation}  \label{intG}
G_{\sigma}(i\omega_n)={\frac{2}{W^2}} \left(\zeta-s\sqrt{\zeta^2-W^2}\right)
\end{equation}
where $s$=sgn[Im$\zeta$].

\begin{eqnarray}  \label{intD.3d}
3d & \qquad\qquad & D(i\omega_n)= {\frac{1}{\xi\omega_1^2}}\left(2\xi+\sqrt{
(\xi-\omega_0)^2-\omega_1^2} -\sqrt{(\xi+\omega_0)^2-\omega_1^2}\right)\\
\label{intD.2d}
2d & \qquad\qquad & D(i\omega_n)= {\frac{1}{4\xi\omega_1}}\ln\left[{\frac{
(\xi+\omega_1)^2-\omega_0^2}{(\xi-\omega_1)^2-\omega_0^2}}\right]
\end{eqnarray}

We consider here a semiclassical treatment of the problem. \ In its most
general form, \ the approach has been described in ref,
\cite{semiclassical} \ and is an application of the saddle point method. \ In
this paper we use a more limited form of this method
that consists of evaluating Eqs. (\ref{defG},\ref{defD}) by a saddle point
technique. It can be viewed as a combination of two separate approximations:
the static approximation (equivalent to the phonon mass $M\rightarrow \infty 
$ limit) and a saddle point analysis of the EDMFT equations in the static
approximation.

The approach of treating the collective excitations as classical, while the
electrons are treated fully quantum mechanically, goes back to the Hubbard
approximation.\cite{Hubbard1964} It was pointed out that a static
approximation of the impurity model coupled with the \ DMFT self consistency
\ conditions indeed gives a solution closely related to Hubbard's.\cite
{Rozenberg92} 
This approach has been used extensively in refs \cite
{Freericks1993,Millis1996} in DMFT studies of the Holstein model. 
From the DMFT studies of the Mott transition, 
\cite{Georges1996} we know that this approach becomes insufficient in the
correlated metallic regime at very low temperatures, where a quasiparticle
feature \ forms in addition to the spectral features produced in the
semiclassical approximation. It is worth pointing out, that improvements
of the static or of the saddle point approximation,\cite{semiclassical}
will not remedy this shortcoming, which requires
a non perturbative resummation of instanton events. Still, we show here that
this simple analysis is able to reproduce all the trends
of the solution of the EDMFT equations by the more expensive QMC\ method. 
\cite{Motome2000}

\bigskip

\bigskip

The EDMFT equations \ in the static approximation Eqs. (\ref{defG},\ref{defD})
reduce to:

\begin{equation}  \label{G}
G(i\omega_n)=\int dx P(x){\frac{1}{G_0^{-1}(i\omega_n)+\lambda x}}
\end{equation}

\begin{equation}  \label{D}
D=-\beta\int dx P(x)x^2
\end{equation}
where

\begin{equation}  \label{p}
P(x)={\frac{1}{N}}\exp\left(g\sum\limits_{n\ge0}
\ln\left(1-G_0(i\omega_n)^{2}\lambda^2 x^2\right)- {\frac{\beta}{2}}
D_0^{-1}x^2\right)
\end{equation}

Eqs. (\ref{G}-\ref{p}) have to be solved together with Eqs. (\ref{selfcons.G},
\ref{selfcons.D}). In the static limit only the zero phonon frequency survives,
so we drop frequency index for the phonon correlation functions $D_0$ and $D$
. In Eqs. (\ref{G}-\ref{p}) and everywhere below we consider $x$ being the
phonon field amplitude, it is related to its Fourier transform as $x=\beta^{-
{\frac{1}{2}}}x_{\omega_m=0}$. We consider no symmetry breaking in the
electron spin channel, so we dropped the spin index; factor $g$ (equal 2 for
spin one-half) in the Eq. (\ref{p}) appears from trace over the spin index. $N$
normalizes $P(x)$ to unity. $P(x)$ is the probability distribution function
of the phonon field amplitude $x$.

We now evaluate Eqs. (\ref{G}-\ref{p}) in the saddle point approximation in
the variable $x$. There are two limits, weak and strong coupling. In the
weak coupling the saddle point is at $x=0$, and in the strong coupling there
are two equivalent saddle points at $x=\pm x_0\ne0$. Deriving the saddle
point equations we explicitly use semicircular electron DOS, Eqs. (\ref{el.dos}
,\ref{intG}). The relation between the bare and full Green's functions is
especially simple in this case: 
\begin{equation}  \label{G0}
G_0(i\omega_n)^{-1}=i\omega_n-t^2G(i\omega_n)
\end{equation}
where $t=W/2$. Everywhere below in the paper energy is measured in units of $
t$. In this paper we restrict ourselves to the particle-hole symmetric case.

In the weak coupling regime in the saddle point approximation, which
includes Gaussian fluctuations of $x$ around zero, semiclassical EDMFT
equations Eqs. (\ref{G}-\ref{p}) read:

\begin{equation}  \label{wcspg}
\tilde G(\tilde G+\omega)^3-(\tilde G+\omega)^2+\alpha^2=0
\end{equation}

\begin{equation}  \label{wcspd}
D_0^{-1}-D^{-1}=-T\sum\limits_{n\ge0} {\frac{2g\lambda^2}{(\tilde G
+\omega_n)^2}}
\end{equation}
where $\tilde G=iG(i\omega_n)$ and $\alpha^2=\lambda^2|D|T=
-\lambda^2 T \int d\epsilon\rho_{ph}(\epsilon)
[D^{-1}-D_0^{-1}-\epsilon^2]^{-1}$, so $\alpha^2$ is
solved for the phonon self energy thus making the system 
of the saddle point equations closed.

In the strong coupling regime we consider two saddle points $x=\pm x_0$. We
discard fluctuations around these points (so $|D|=\beta x_0^2$), since
nontrivial information is contained in the fact that we have two saddle points,
and not in the Gaussian fluctuations, like it was in the case of weak
coupling. EDMFT equations Eqs. (\ref{G}-\ref{p}) now read:

\begin{equation}  \label{scspg}
\tilde G(\tilde G+\omega)^2-(\tilde G+\omega)+\tilde G\alpha^2=0
\end{equation}

\begin{equation}  \label{scspd}
D_0^{-1}=-T\sum\limits_{n\ge0} {\frac{2g\lambda^2}{(\tilde G
+\omega_n)^2+\lambda^2TD}}
\end{equation}

Weak coupling equations Eqs. (\ref{wcspg},\ref{wcspd}) are a saddle point
expansion up to the first order in small parameter $\lambda^2DT$, and strong
coupling equations Eqs. (\ref{scspg},\ref{scspd}) - up to the first order in
large parameter $\lambda^2D/T$. These equations have overlapped regions of
applicability, provided $T\ll1$. This allows us to combine weak and strong
coupling equations into a unique set of equations, controlled by the small
parameter $T$:

\begin{equation}  \label{spg}
\tilde G(\tilde G+\omega)^2-(\tilde G+\omega)+\tilde G\alpha^2=0
\end{equation}

\begin{equation}
D_{0}^{-1}-D^{-1}=-2g\lambda ^{2}T\sum\limits_{n\geq 0}{\frac{\tilde{G}}{
\tilde{G}+\omega _{n}}}  \label{spd}
\end{equation}
These are our final semiclassical EDMFT equations. They are exact in the
limit $MT^{2}\gg \omega _{0}^{2}$, $T\ll 1$. For 3d and 2d phonons Eqs. (\ref
{spg},\ref{spd}) have to be solved together with Eq. (\ref{intD.3d},\ref
{intD.2d}), where $\xi ^{2}=D^{-1}-D_{0}^{-1}$. Saddle point equations Eqs. 
(\ref{spg},\ref{spd}) are very simple, they can be solved for $D$ and $D_{0}$
with minimal numerical efforts. Left hand side (lhs) part of 
Eq. (\ref{spg}) is a third degree
polynomial, so electron Green's function can be written as an elementary
function determined by a single parameter $\alpha ^{2}$ 
which is a function of phonon self energy and bare parameters of the model. 

In the limits of small
and large $\alpha ^{2}$ (or $\lambda $) Eqs. (\ref{spg},\ref{spd}) are solved
completely for self energies:

$\alpha\ll1$

\begin{equation}  \label{else.wc}
\Sigma_{el}(i\omega_n)= \left(-{\frac{\omega_n}{2}}+\sqrt{1+({\frac{\omega_n
}{2}})^2}\right) \alpha^2
\end{equation}

\begin{equation}
\Sigma _{ph}=-{\frac{4g}{3\pi }}\lambda ^{2}  \label{phse.wc}
\end{equation}
Moreover, in the dispersionless case 
$\alpha ^{2}=\lambda ^{2}T/(\omega _{0}^{2}-{
\frac{4g}{3\pi }}\lambda ^{2})$. This expression is valid
everywhere except for the small region $\Delta \lambda \sim \omega _{0}T$
below $\lambda _{c}\sim \omega _{0}$.
We consider here a disordered phase solution. In $d=3$ the disorder solution
becomes unstable at $\lambda \sim \omega _{0}-\omega _{1}$ while it remains
stable for all coupling in $d=2$. \ The self energies in the
strong coupling regime $\alpha \gg 1$ are given by :

\begin{equation}  \label{else.sc}
\Sigma_{el}(i\omega_n)={\frac{\alpha^2}{\omega_n}}
\end{equation}

\begin{equation}
\Sigma _{ph}=-{\frac{g\lambda ^{2}}{2\alpha }}  \label{phse.sc}
\end{equation}
In the strong coupling the phonon field distribution function is split in
two peaks. The peak separation is $2x_{0}$, where $x_{0}=-g\lambda
D_{0}/2 $, $D=-\beta x_{0}^{2}$. In the dispersionless case 
$\alpha ^{2}=({\frac{g}{2}}\lambda ^{2}/\omega _{0}^{2})^{2}$, this is
valid when $\lambda \gg \omega _{0}$. 
This is completely similar to the previous analysis.\cite{Millis1996}

In $d=3$ the\ instability to the ordered phase occurs already at small 
$\omega_{1}\sim 
{\frac{\omega _{0}^{3}}{\beta g^{2}}}{\frac{1}{\lambda ^{2}}}$, so 
$D_{0}$ stays practically unrenormalized.

In $d=2$ at $\omega _{1}\sim {\frac{\omega _{0}^{3}}{\beta g^{2}}}{\frac{1}{
\lambda ^{2}}}$ the system enters a regime when \ the phonon \ energy is
exponentially small:

\begin{equation}  \label{exp.soft}
\Sigma_{ph}-(\omega_0-\omega_1)^2\approx2\omega_1\exp{\left[ -\beta g {\frac{
\omega_1}{\omega_0^3}}\lambda^2\right]}
\end{equation}

In the limit $T\to0$ one readily obtains the polaron formation condition, which
happens at intermediate ($\lambda_c\sim\omega_0$) coupling: 
\begin{equation}  \label{doublepeak}
-{\frac{4g}{3\pi}}\lambda_c^2D_0=1
\end{equation}
where $D_0=\omega_0^{-2}$ in the dispersionless case, but has to be found
numerically for interacting phonons.

\section{3D phonons}

\label{sec_3d}

In this section we compare our semiclassical solution to the exact QMC
results.\cite{Motome2000} 
The saddle point equations we derived are exact when 
$(2\pi T)^{2}M\omega _{0}^{-2}\gg 1$ and $Tt^{-1}\ll 1$.
The QMC results\cite{Motome2000} however, 
were obtained for $(2\pi T)^{2}M\omega _{0}^{-2}\approx
2.5$ and $Tt^{-1}\approx 0.13$. We want to show, that even in these cases
when the parameters controlling the saddle point equations are relatively
close to $1$, \ the semiclassical solution,
even without including the refinements outlined in ref \cite{semiclassical}
not only captures all the qualitative trends of the exact solution, but in
many instances is quantitatively close to it.

We study \ the\ case of 3-dimensional phonons. We use the same parameters as
in the ref\cite{Motome2000}: inverse temperature $\beta =8$, the phonon band
is centered at $\omega _{0}=.5$, electrons have double spin degeneracy $g=2$
and hopping amplitude $t=1$, phonon mass $M=1$. The electron band is half
filled. To model 3d phonons semicircular DOS Eq. (\ref{ph.3d.dos}) is used. \
We present the solution of Eqs. (\ref{spg},\ref{spd}) and Eq. (\ref{intD.3d}).

In every figure in this section we plot both our and QMC curves. Our results
are plotted using solid or dashed lines only, without symbols. QMC graphs
are presented using dotted lines and always with symbols.

\begin{figure}
\epsfxsize=4.5 truein
\epsfbox{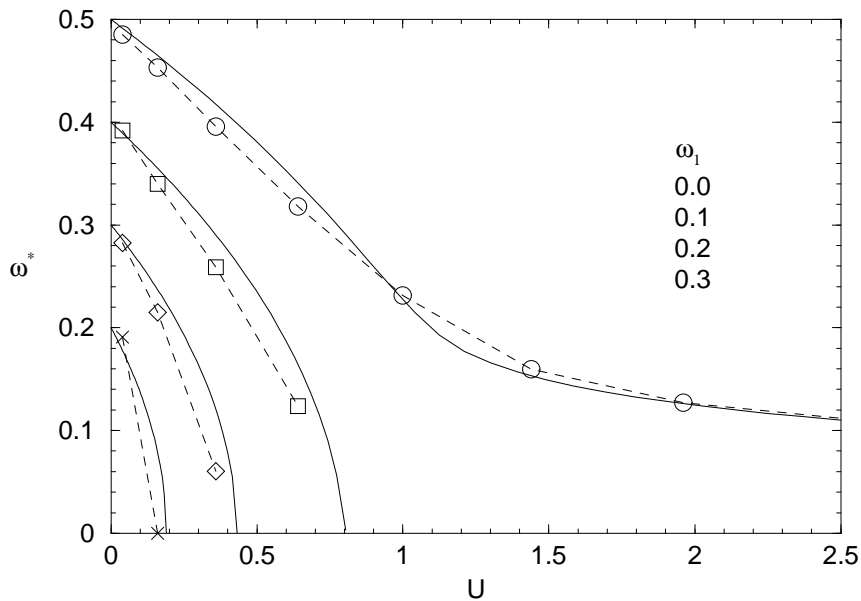}
\caption{$d=3$. Effective phonon frequency $\protect\omega^*$ as a function
of $\protect\lambda^2/\protect\omega_0^2$ at $\protect\omega_1={0.0, 0.1,
0.2, 0.3}$. Comparison to QMC.}
\label{soft.3d}
\end{figure}

A local instability, starting from the disordered phase, 
takes place within  EDMFT when as discussed in ref\cite{Motome2000}  
the  effective phonon frequency $\omega ^{\ast }=
\sqrt{(\omega _{0}-\omega _{1})^{2}+\Pi }$, given by the pole in the phonon
Green's function, becomes equal to zero. The phonon mode softening for
different values of the phonon dispersion are shown in Fig.\ref{soft.3d}. $
\omega ^{\ast }$ is plotted versus the quantity characterizing the effective
interaction: $U=\lambda ^{2}/\omega _{0}^{2}$. The effective
electron-electron interaction, mediated by phonons is given by $
U_{eff}=\lambda ^{2}D_{0}$ \ evaluated at zero frequency .
When the phonon dispersion vanishes $U_{eff}=U$. The upper curve 
in Fig.\ref{soft.3d} corresponds to the dispersionless case.

On the other hand,  we find this is not the 
best way to detect an instability to an ordered phase
and we discuss in Appendix \ref{stability} an alternative way to
compute within EDMFT 
the phonon self energy which retains momentum dependence. 

The equations in section \ref{sec_spe} do not include 
possibility of the phonon
field symmetry breaking. They need to be modified to describe long range
ordering,\cite{chitra2} and we implement this in section \ref{tc}. 
A well known property of mean field theories is that they allow the
analytic continuation of mean field solutions beyond the parameter regime
where they are stable. This was very fruitful in the understanding of the
paramagnetic Mott insulating phase which is unstable to ferromagnetism. 
\cite{Georges1996} \ \ As was done by QMC \ in ref,\cite{Motome2000} we
study the continuation of the EDMFT equations beyond paramagnetic phase. It
may hopefully be understood as a metastable phase. This requires some care
since the instability to a charge ordered phase is signaled by a singularity
appearing in the integrand in Eq. (\ref{enintD}) and this instability causes $
D$ \ to acquire an imaginary part. \ As in ref,\cite{Motome2000} we take
the principal part of the integrand, the imaginary part of $D_0$, 
being equal to zero in every numerical iteration loop, which allows us 
to compare our results with the results of the QMC method. 

\bigskip

\subsection{Weak coupling}

\begin{figure}
\epsfxsize=6.0 truein
\epsfbox{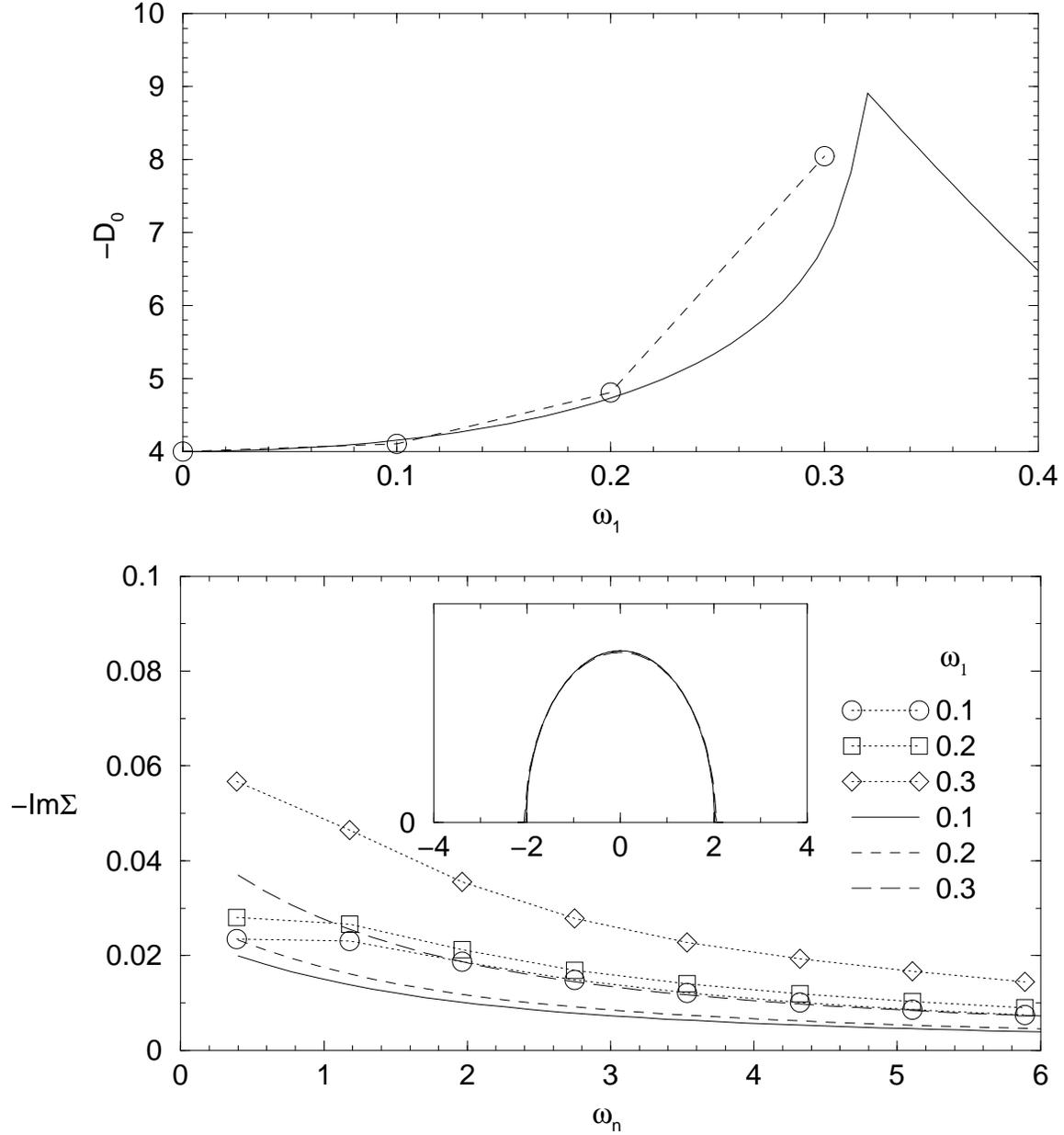}
\caption{$d=3$. Weak coupling $\protect\lambda=.2$. Comparison to QMC. a)
The bare phonon Green's function. b) The imaginary part of the electron self
energy with the spectral function in the inset. 
$\protect\omega_1={0.1, 0.2, 0.3}$.}
\label{d0.wc.3d}
\label{se.wc.3d}
\end{figure}

The finite dispersion treated within DMFT renormalizes $D_{0}$ (see Fig.\ref
{d0.wc.3d}(a)). Since the effective electron electron interaction is
proportional to $D_{0}$, the electron self energy is enhanced as well (see
Fig.\ref{se.wc.3d}(b)). While the features of the exact solution are
qualitatively well reproduced in the semiclassical approach, it lacks
quantitative agreement. The weak coupling is the worst case. The quantitative
agreement is better for intermediate and strong coupling.

\subsection{Strong coupling}

\begin{figure}
\epsfxsize=6.0 truein
\epsfbox{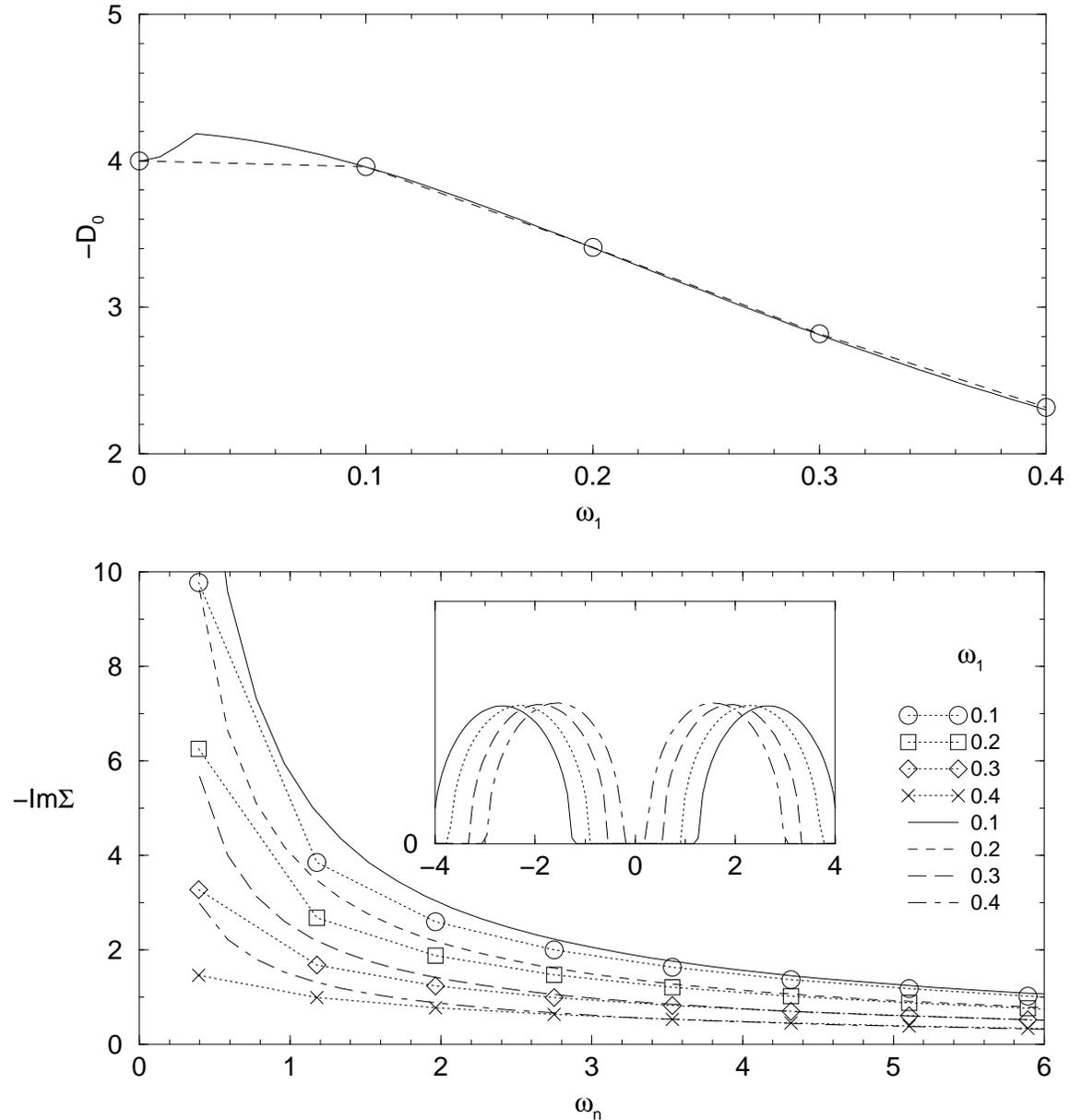}
\caption{$d=3$. Strong coupling. $\protect\lambda=.8$. a) Bare phonon Green's
function. b) Imaginary part of the electron self energy 
with the spectral function in the inset. $\protect\omega_1= {
0.1, 0.2, 0.3, 0.4}$}
\label{d0.sc.3d}
\label{se.sc.3d}
\end{figure}

In the strong coupling regime as dispersion increases, $D_0$ renormalizes
downward (see Fig.\ref{d0.sc.3d}(a)), together with the electron self energy
(see Fig.\ref{se.sc.3d}(b)). For the strong coupling the quantitative
agreement with QMC is very good.

\subsection{Intermediate coupling}

\begin{figure}
\epsfxsize=6.0 truein
\epsfbox{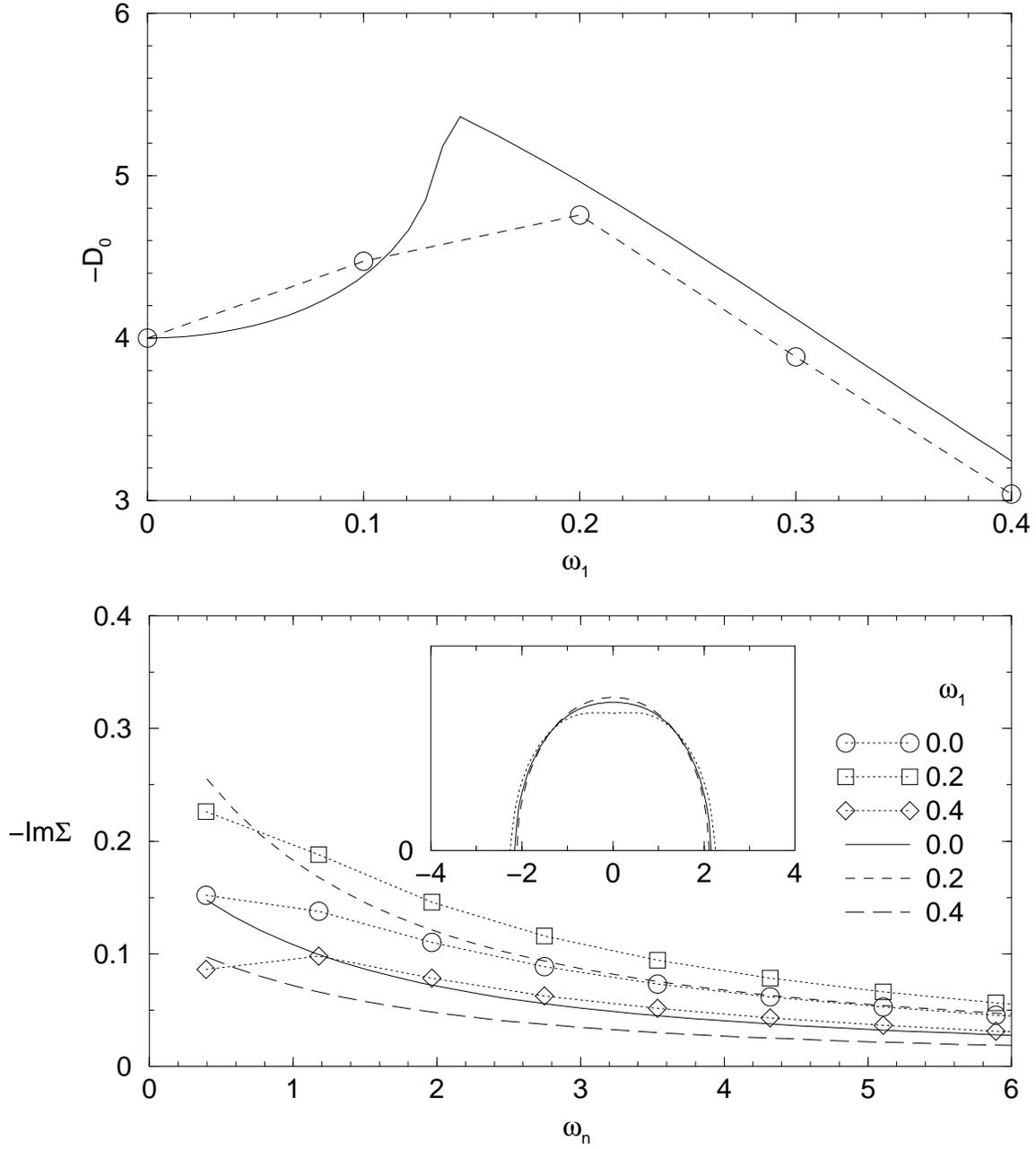}
\caption{$d=3$. Intermediate coupling $\protect\lambda=.4$. Comparison to
QMC. a) Bare phonon Green's function. b) Imaginary part of the electron self
energy with the spectral function in the inset.
 $\protect\omega_1={0.0, 0.2, 0.4}$.}
\label{d0.ic.3d}
\label{se.ic.3d}
\end{figure}

At intermediate coupling system is in a crossover between weak and strong
coupling regimes. As $\omega_1$ increases, effective electron-electron
interaction first becomes stronger, $D_0$ and electron self energy
increases, like at weak coupling. At $\omega^*=0$ behavior changes on
reverse, the picture is similar to strong coupling case. This is illustrated in
Fig.\ref{d0.ic.3d}(a) and Fig.\ref{se.ic.3d}(b).

\section{2D phonons}

\label{sec_2d}

In the previous section we calculated various functions at different
parameters in the 3d case. \ The saddle point \ approximation is exact in the
limit of infinite mass $M$ and zero temperature $T$. At finite $M$ and $T$
the applicability of the method in a wide region of parameters was
established in the previous section by comparison to QMC data. 
In this section we study 2d phonon case (Eqs. (\ref{spg},\ref{spd}) 
and Eq. (\ref{intD.2d})) in the same range of parameters.

Unlike the 3d case, the 2d disorder solution is locally stable, and we focus
on this solution in this section. As we will show in section \ref{tc}
EDMF in dimensions \ $d<4$ , gives rise to a first order transition at a
critical coupling strength. We study the disordered state solution 
continued along the second order transition branch, 
which is skipped in the first order transition.

In the EDMFT approach, $d=2$, appears as a lower critical dimension
for finite temperature second order transition. This result \ describes
accurately the situation with order parameters posessing a continuous
symmetry, but it is a spurious consequence of the inability of a local
approximation to generate spatial anomalous dimensions in the cases where
order breaks a discrete symmetry.

\begin{figure}
\epsfxsize=5.3 truein
\epsfbox{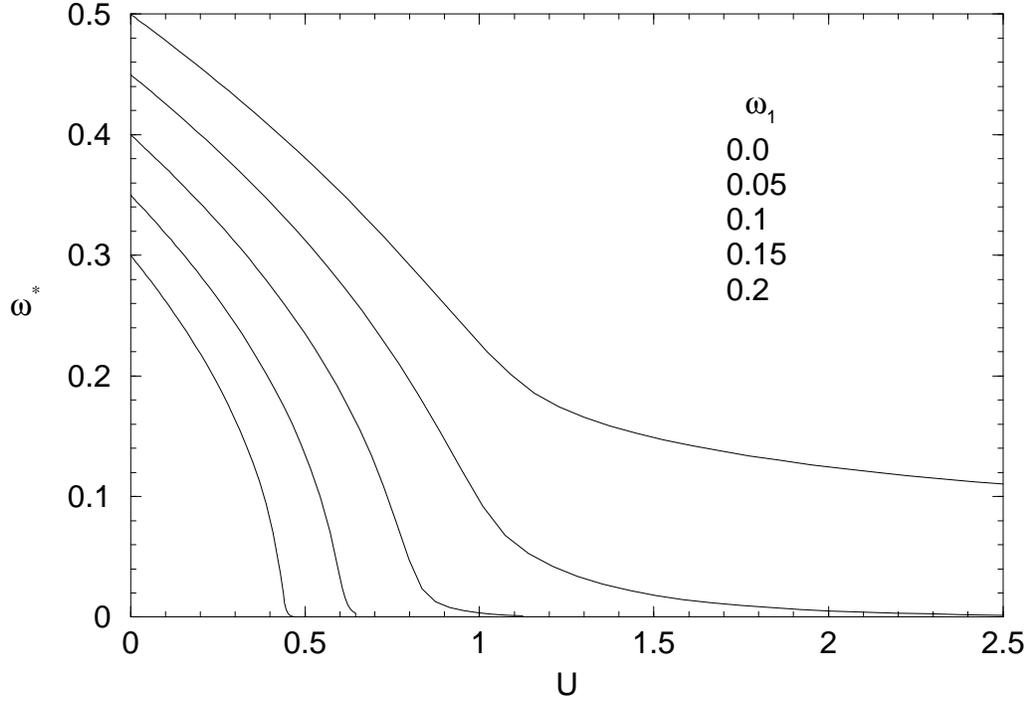}
\caption{$d=2$. Effective phonon frequency $\protect\omega^*$ as a function
of $\protect\lambda^2/\protect\omega_0^2$ at $\protect\omega_1= {0.0, 0.05,
0.1, 0.15, 0.2}$}
\label{soft.2d}
\end{figure}

First we illustrate the exponential softening of the collective mode : in
Fig.\ref{soft.2d} we plot effective frequency $\omega ^{\ast }$ versus $U$
. The Fig.\ref{soft.2d} should be compared to Fig.\ref{soft.3d} (3d
case). In the latter the curves hit $U$ axis, what implies second order
transition. In Fig.\ref{soft.2d} the curves rather gradually approach $U$
axis, never crossing it.

Phonons generate effective electron-electron interaction $\sim \lambda^2 D_0$
, so we are especially interested in $D_0$ behavior. We investigate 2d
system for similar sets of parameters as we did in 3d case.

We obtained the following plots in weak, intermediate and strong coupling
regimes:

\begin{figure}
\epsfxsize=5.5 truein
\epsfbox{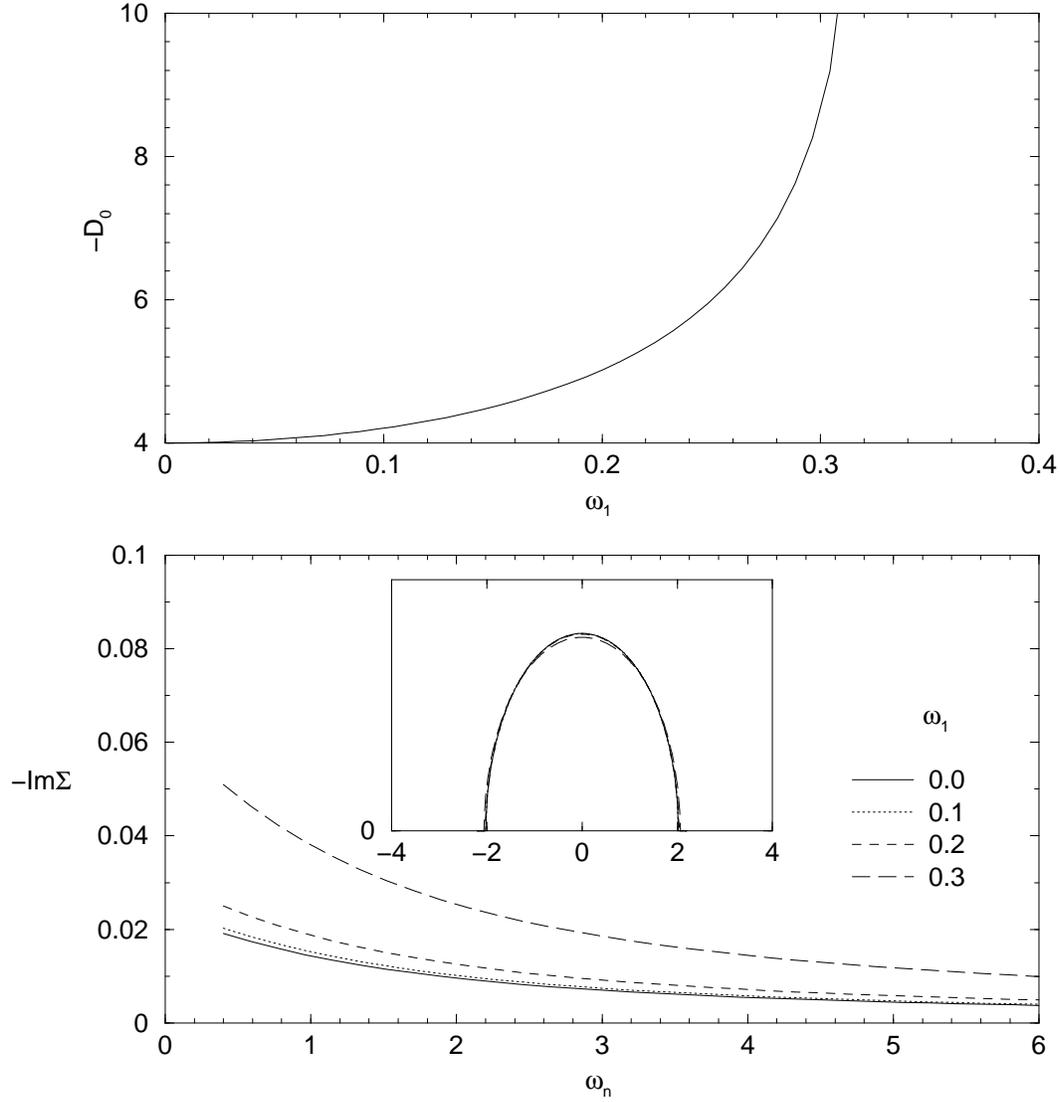}
\caption{$d=2$. Weak coupling. $\protect\lambda=.2$. a) The bare phonon
Green's function. b) The imaginary part of the electron self energy 
with the spectral function in the inset. $\protect
\omega_1={0.0, 0.1, 0.2, 0.3}$.}
\label{se.wc.2d}
\end{figure}

\begin{figure}
\epsfxsize=5.5 truein
\epsfbox{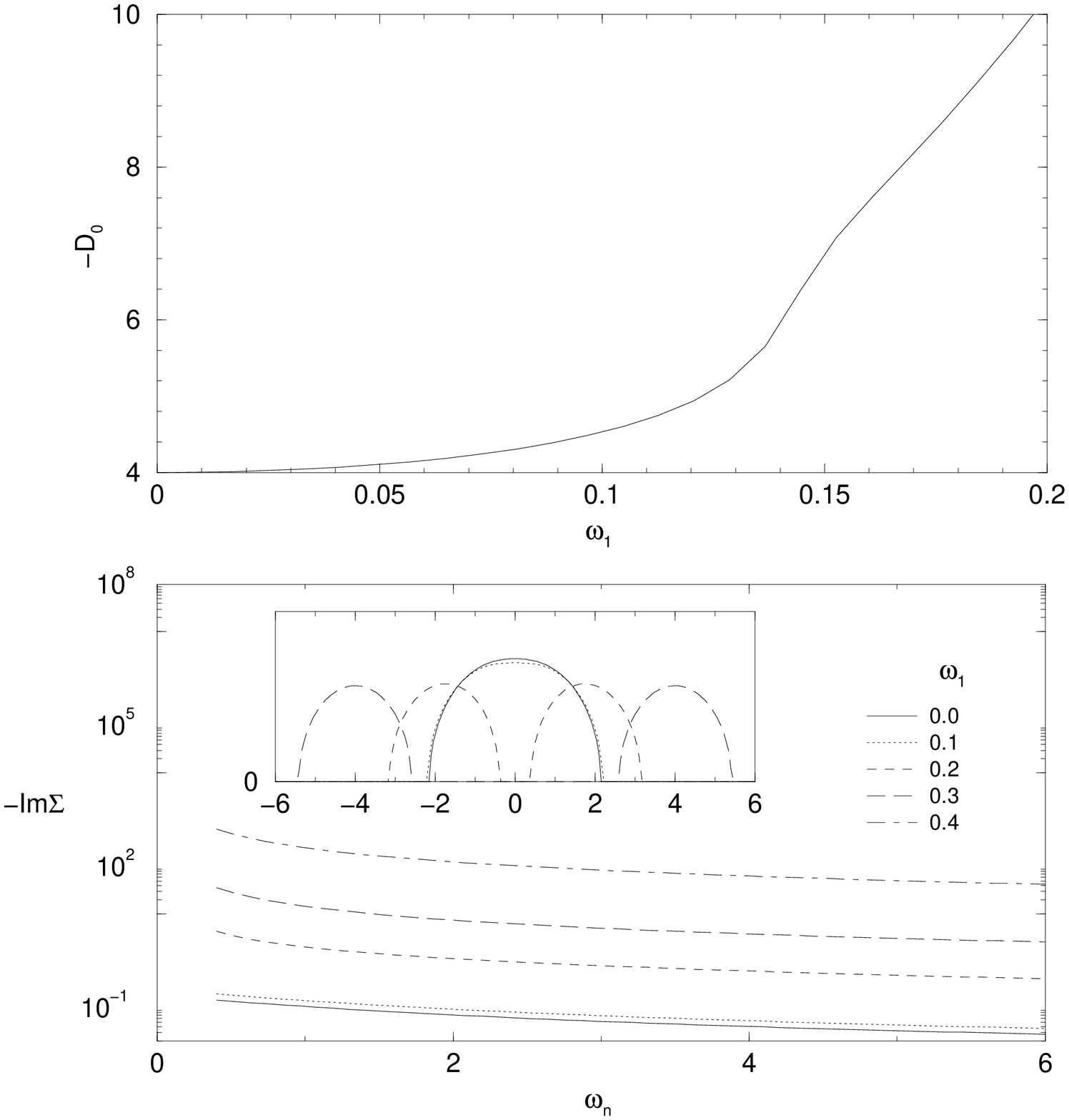}
\caption{$d=2$. Intermediate coupling. $\protect\lambda=.4$. a) The bare
phonon Green's function. b) The imaginary part of the electron self energy 
with the spectral function in the inset. 
$\protect\omega_1={0.0, 0.1, 0.2, 0.3, 0.4}$}
\label{se.ic.2d}
\end{figure}

\begin{figure}
\epsfxsize=5.5 truein
\epsfbox{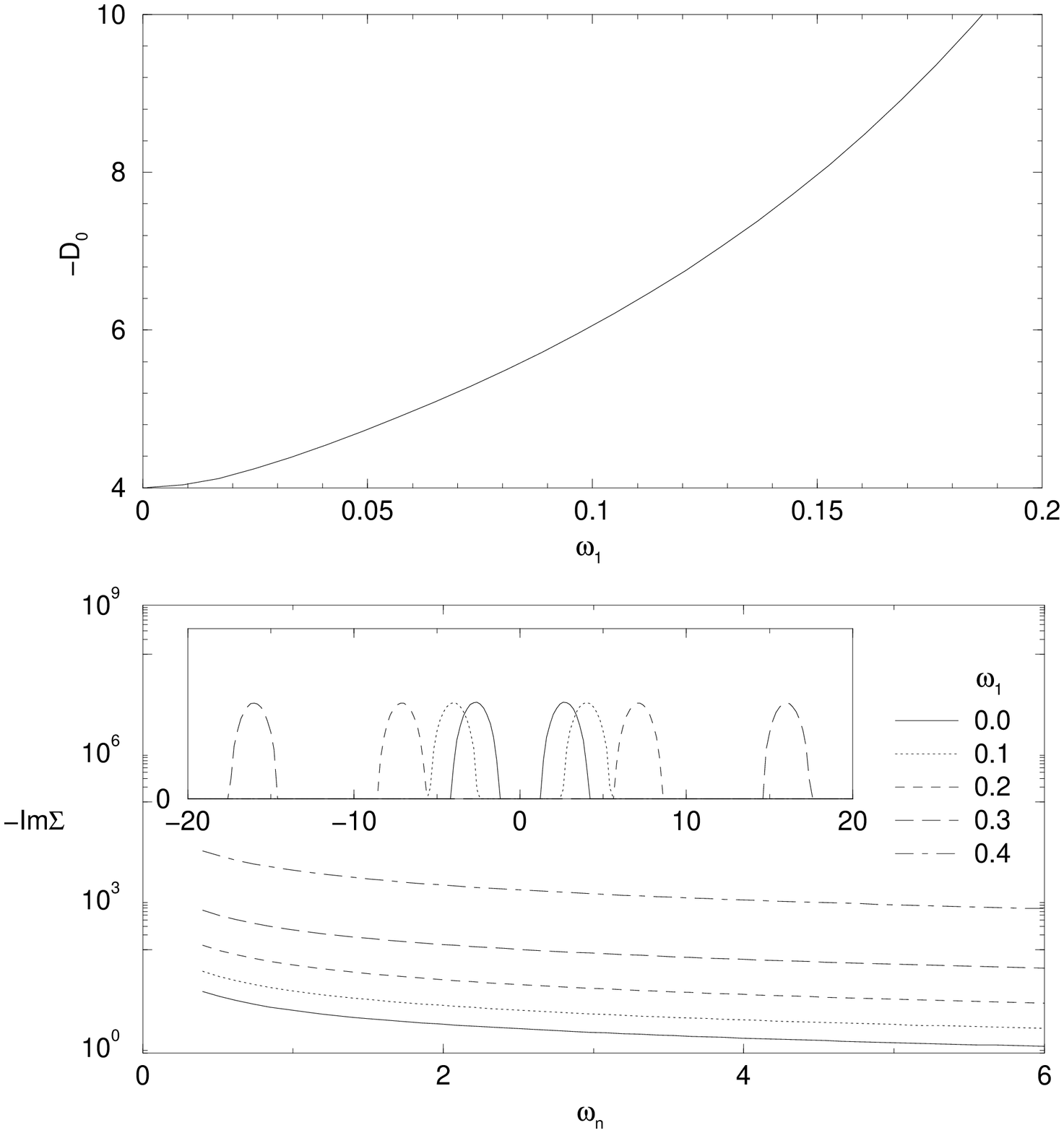}
\caption{$d=2$. Strong coupling. $\protect\lambda=.8$. a) The bare phonon
Green's function. b) The imaginary part of the electron self energy 
with the spectral function in the inset. $\protect
\omega_1=0.0, 0.1, 0.2, 0.3, 0.4$}
\label{se.sc.2d}
\end{figure}

\begin{figure}
\epsfxsize=4.7 truein
\epsfbox{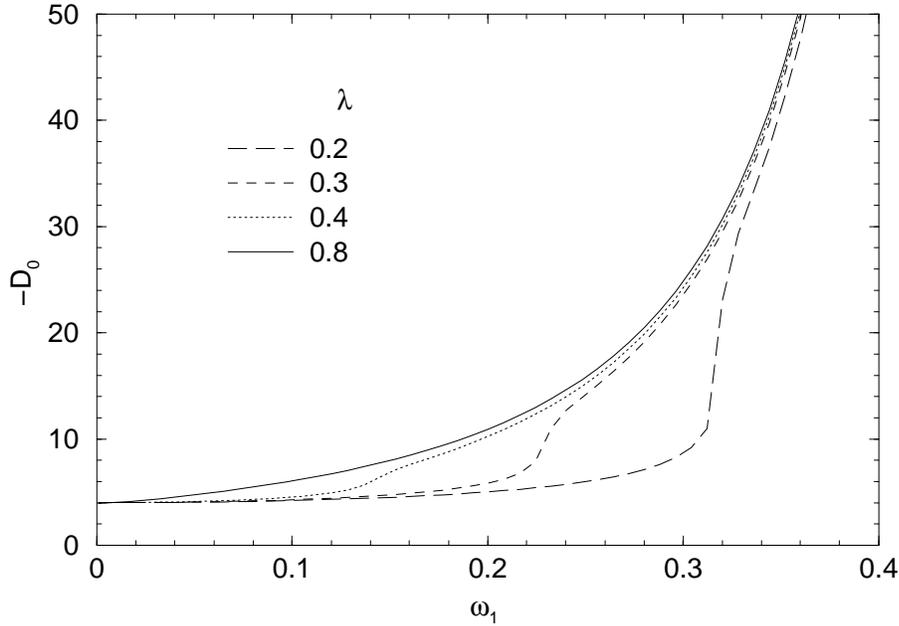}
\caption{$d=2$. Bare phonon Green's function. $\protect\lambda=.2,.3,.4,.8$}
\label{d0.2d}
\end{figure}

The behavior of 2d system is very similar to that of 3d system before the
energy of the \ phonon mode vanishes. In all cases $D_{0}$ gets
renormalized, as \ the dispersion and consequently the effective interaction
increases. The electron self energy enhances correspondingly. The only
difference is the rate of $D_{0}$ renormalization in the weak, strong and
intermediate coupling. $D_{0}$ renormalizes faster at larger $\lambda $ (see
Fig.\ref{d0.2d}), since electrons are stronger coupled to phonons.

\section{Ordered Phase and Critical Temperature}

\label{tc}

We now turn to the generalization of the EDMFT equations to the ordered
phase.\cite{chitra2} For simplicity we will consider a classical model.
This is justified, since in the semiclassical limit we can always integrate
out the electrons, reducing EDMFT equations to classical mean field
equations. 
For instance tracing out the electrons and performing a static approximation
in the electron-phonon \ field leaves us with an action of the form (
neglecting terms of order $\phi ^{6}$ and higher):

\begin{equation}
S[\phi]=\sum\limits_i{\frac{r}{2}}\phi_i^2+{\frac{U}{4}}\phi_i^4
-\sum\limits_{ij}\phi_i {\frac{J_{ij}}{2}}\phi_j=
S_{loc}[\phi]-\sum\limits_{ij}\phi_i {\frac{J_{ij}}{2}}\phi_j
\label{class.act}
\end{equation}

\begin{figure}
\vspace{.2in}
\epsfxsize=6 truein
\epsfbox{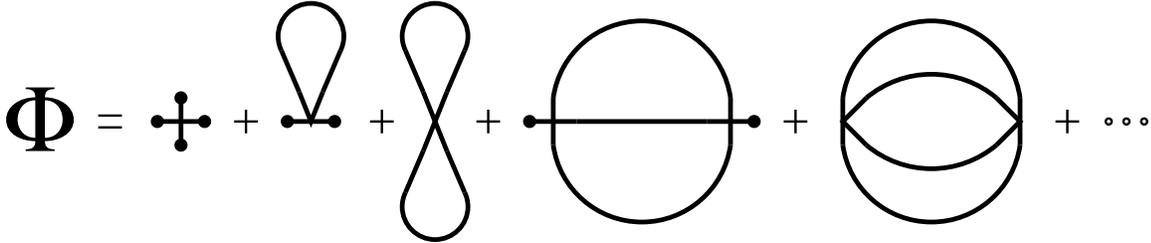}
\vspace{.1in}
\caption{Diagrammatic expansion of $\Phi$ up to the first two orders in $U$}
\label{phi.diagrams}
\end{figure}

To extend \ the EDMFT\ approach to the ordered phase it is useful to \ write
down the Baym Kadanoff functional for the action, 
\begin{equation}
\label{bkfunctional}
\Gamma \lbrack m,D]=-{\frac{1}{2}}{\rm Tr}\log {D}+{\frac{1}{2}}{\rm Tr}
D_{0}^{-1}D+{\frac{1}{2}}mD_{0}^{-1}m+\Phi \lbrack m,D]
\end{equation}
$\Phi $ is a sum of all two particle irreducible diagrams constructed from
phonon Green's functions $D$, phonon field expectation value $m$ and four
legged interaction vertex $-3!U$. We could also say that 
$\Phi $ is a sum of all two particle irreducible diagrams constructed from
phonon Green's functions $D$ and 
four, three, two legged vertices plus the
first diagram shown in figure (\ref{phi.diagrams}), \ which contains no
propagators. The vertices yield factors of $-3!U$, $-3!Um$, $-3Um^{2}$ and 
$-{\frac{1}{4}}Um^{4}$ for four, three, two and zero legged vertices
respectively. 
Each diagram in $\Phi $ has an extra $-1$ factor.

 In the Fig.\ref{phi.diagrams} we drew first and second order 
(in $U$) diagrams
entering $\Phi $. \ The extended DMFT equations in the ordered phase are
derived by making the local approximation on $\Phi$ in the 
Baym Kadanoff functional
and solving the stationary conditions for the magnetization and the local
propagator resulting from the stationarity of Eq. (\ref{bkfunctional})
after this
local approximation is made. In the local approximation the leading terms
in a perturbative expansion in the quartic coupling are given by $\Phi =
{\frac{1}{4}}Um^{4}+{\frac{3}{2}}UDm^{2}+{\frac{3}{4}}UD^{2}-3U^{2}D^{3}m^{2}
-{\frac{3}{4}}U^{2}D^{4}+...$

\begin{figure}
\vspace{.2in}
\epsfxsize=6 truein
\epsfbox{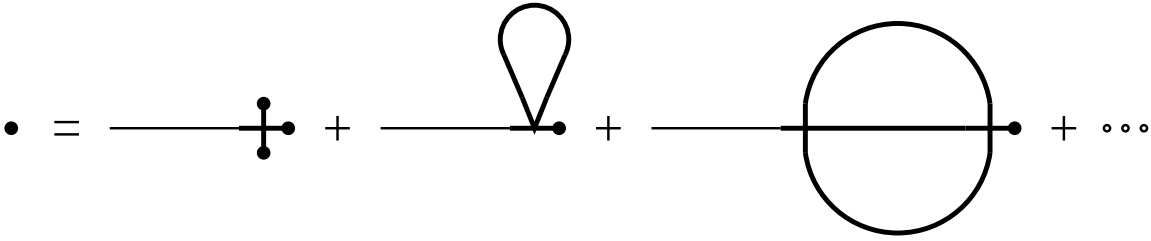}
\vspace{.6in}
\caption{Diagrammatic expansion of $m$ up to the first two orders in $U$.
A thin line is the free phonon propagator $D_0$, a thick line is the
full phonon propagator. A full dot stands for $m$.}
\label{mexpansion}
\end{figure}

Stationarity of the functional \ \ in Eq. (\ref{bkfunctional}) 
would give exact equations for $D$
and $m$. In the local approximation these equations reduce 
to EDMFT equations in
zero magnetization and therefore generalize those to the 
ordered phase.\cite{chitra2} They are given by

\begin{eqnarray}
&m (r-J_{q=0})+{\frac{\delta \Phi }{\delta m}} = 0  \nonumber \\
&D = \sum\limits_{q}[r-J_{q}+2{\frac{\delta \Phi }{\delta D}}]^{-1}
\label{edmfteqn}
\end{eqnarray}
where only local graphs are included in $\Phi $ .
Diagram series equivalent to the first equation are shown 
in Fig.\ref{mexpansion}.

For the practical solution of the EDMFT Eqs. (\ref{edmfteqn}) , it is useful
to follow the dynamical mean field procedure of introducing
an impurity local effective action \cite{gk92,Georges1996} to sum up the
graphs generated by the functional for $\Phi $ and \ its functional
derivatives $\delta \Phi /\delta D$ and $\delta \Phi /\delta m$ in terms
of the cavity \ fields $h$ and $\Delta $ . \ The effective action

\begin{equation}
S_{EDMFT}[\phi]=S_{loc}[\phi]-h\phi-{\frac{\Delta}{2}}\phi^2
\label{edmftact}
\end{equation}
generates the \ correct local quantities provided that the Weiss fields $h$
and $\Delta $ \ are chosen to obey the EDMFT\ self consistency conditions:

\begin{equation}
r-\Delta-D^{-1}=-2{\frac{\delta\Phi}{\delta D}}  \label{del.loc}
\end{equation}

\begin{equation}
h=-m(r-\Delta )-{\frac{\delta \Phi }{\delta m}}  \label{h.loc}
\end{equation}

Eqs. (\ref{edmfteqn}) and Eqs. (\ref{del.loc},\ref{h.loc}) are a closed set of
EDMFT equations, describing both ordered and disordered phase of the
classical system Eq. (\ref{class.act}):

\begin{equation}
D=\sum\limits_q[D^{-1}+\Delta-J_q]^{-1}  \label{edmfteq1}
\end{equation}

\begin{equation}
S_{EDMFT}[\phi]=S_{loc}[\phi]- m\left(J_{q=0}-\Delta\right)\phi -{\frac{
\Delta}{2}}\phi^2  \label{edmfteq2}
\end{equation}

\begin{equation}
m=\langle \phi\rangle_{S_{EDMFT}}  \label{edmfteq3}
\end{equation}

\begin{equation}
D=\langle \phi^2\rangle_{S_{EDMFT}}-m^2  \label{edmfteq4}
\end{equation}

The equations \ above are consistent in describing the transition:
magnetization vanishing in the ordered phase, and divergence of spin
susceptibility across the transition do occur at the same critical
temperature.

When $U\rightarrow \infty $, $r\rightarrow -\infty $, $U/r=-1$ the system,
described by the action Eq. (\ref{class.act}) reduces to a classical Ising
model with spin values $\pm 1$. In this limit the standard Weiss mean field
\ equations 
\begin{equation}
m=\tanh {mJ_{q=0}}  \label{isingmft}
\end{equation}

can be compared with the EDMFT equations which now read:

\begin{eqnarray}
&m=\tanh{m\left(J_{q=0}-\Delta\right)}  \nonumber \\
&1-m^2=\sum\limits_q \left[(1-m^2)^{-1}+\Delta-J_q\right]^{-1}
\label{isingedmft}
\end{eqnarray}

We will also compare the EDMFT\ equations to an extension of mean field
theory due to Bloch \ and Langer (BL):\cite{BlochLanger}

\begin{equation}
M_1=\int\limits_{-\infty}^{+\infty}dx(2\pi G_2)^{-{\frac{1}{2}}} \exp[-{
\frac{(M_1J_{q=0}-x)^2}{2G_2}}]\tanh{x}  \label{bl1}
\end{equation}

\begin{equation}
M_2=\int\limits_{-\infty}^{+\infty}dx(2\pi G_2)^{-{\frac{1}{2}}} \exp[-{
\frac{(M_1J_{q=0}-x)^2}{2G_2}}](\cosh{x})^{-2}  \label{bl2}
\end{equation}

\begin{equation}
G_{2}=\sum\limits_{q}{\frac{J_{q}}{1-J_{q}M_{2}}}  \label{bl3}
\end{equation}
where $M_{1}$ is the magnetization. It can be shown that EDMFT counts
(without overcounting) more terms in diagrammatic expansion of various
physical quantities, like correlation function or free energy, than BL
method does. One can also check that for the classical $\pm1$ spin model
EDMFT gives a better estimate for $T_{c}$ than BL method does.

\begin{figure}
\epsfxsize=5.5 truein
\epsfbox{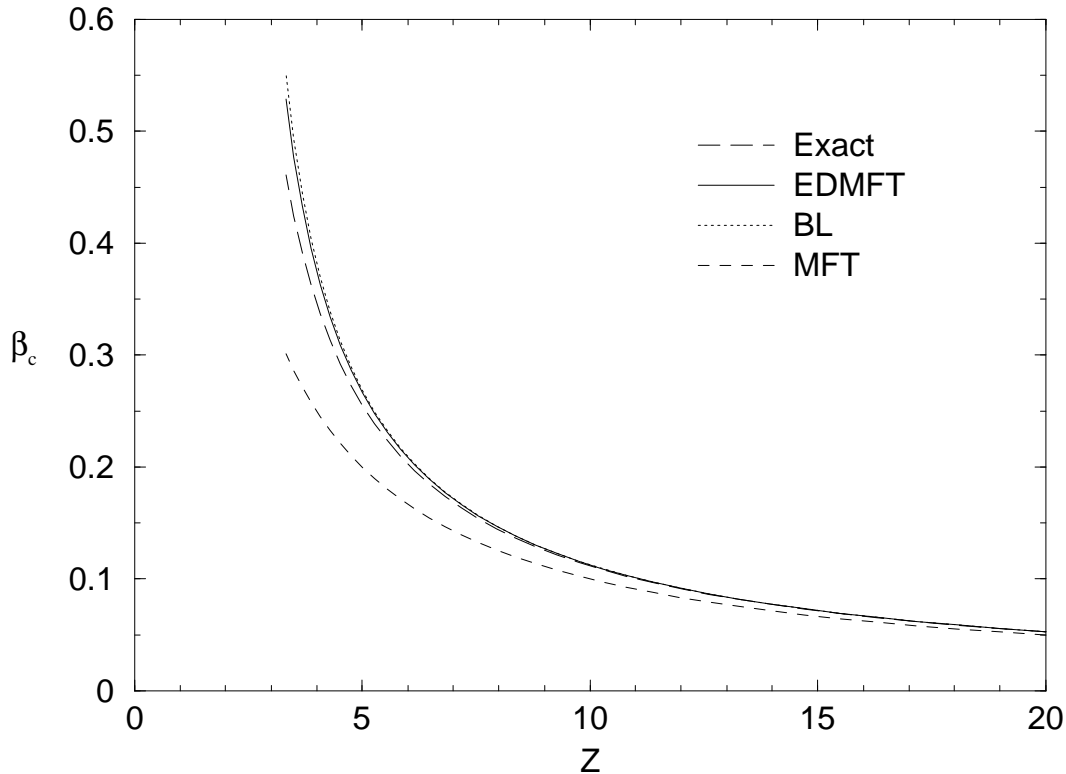}
\caption{Top to bottom: BL, EDMFT, exact solution, MFT; Critical $J_c$
 vs nearest neighbors number $z$}
\label{figtc}
\end{figure}

We computed \ the critical temperature $T_{c}$ for the Ising model on a
Bethe lattice \ with finite coordination following the paramagnetic solution
till it disappears using the different approximations described in this
section. The results are shown in Fig.\ref{figtc}. The EDMFT result shows
significant improvement over MFT, and is slightly better than BL method.
Some more technical details \ comparing the approximation schemes are
relegated to Appendix \ref{ring diagrams}. \ 

\begin{figure}
\epsfxsize=5.5 truein
\epsfbox{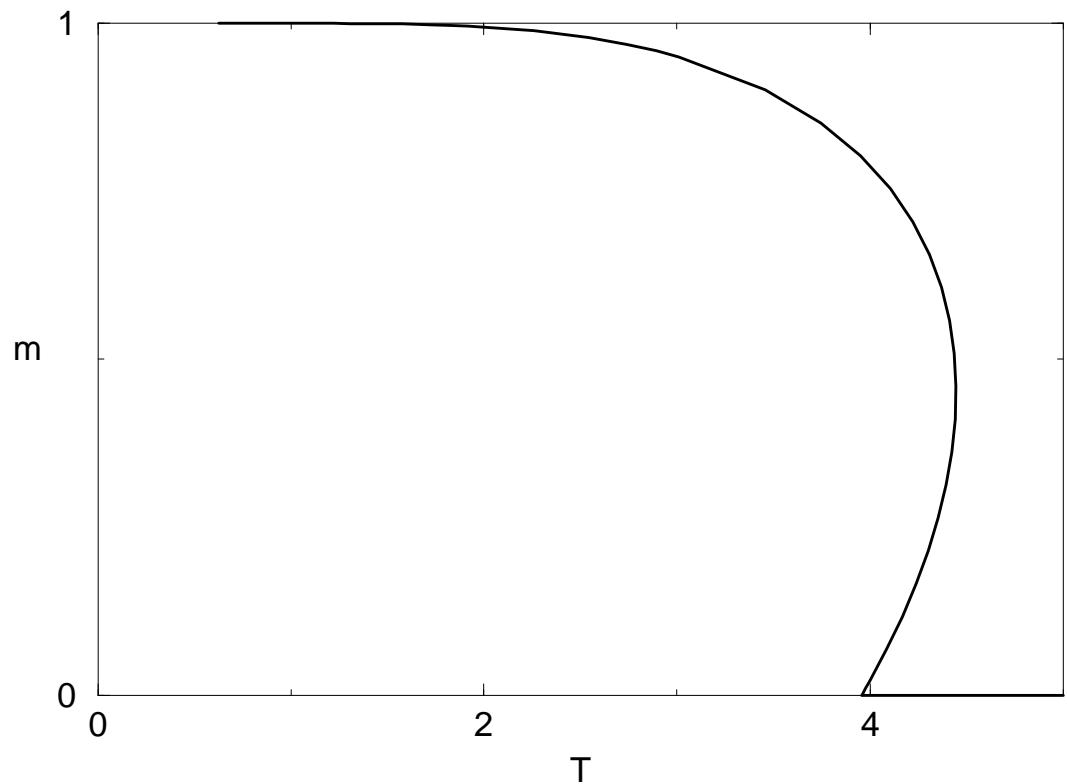}
\caption{Magnetization vs. temperature on a cubic lattice.
There are three branches at $3.96<T<4.45$: $m_3>m_2>m_1=0$, 
$m_1$ and $m_3$ are physical
solutions, while $m_2$ is not. 
Classical $\pm1$ spin Ising model.}
\label{magn3d}
\end{figure}

\begin{figure}
\epsfxsize=5.5 truein
\epsfbox{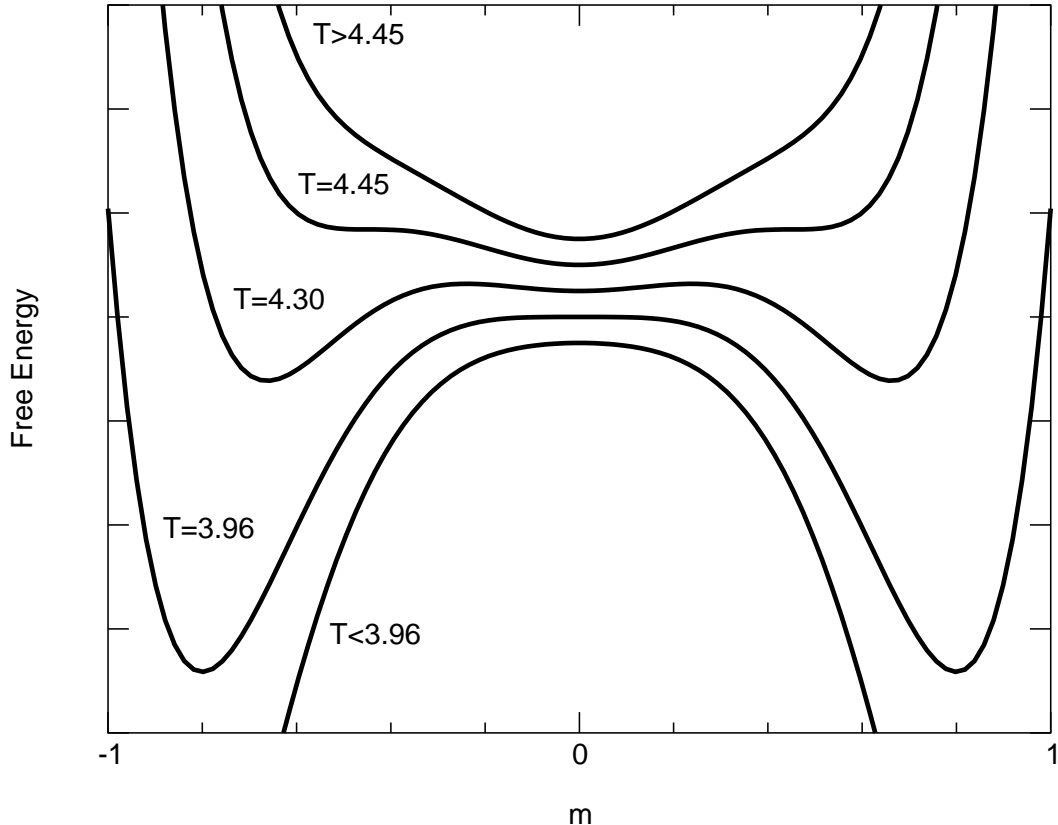}
\vspace{.3in}
\caption{Free energy evolution with $T$. Free energy has a single
minimum ($m=0$) above $T=4.45$; at $T=4.45$ the solution bifurcates
at $m=.45$. There are free extrema at $3.96<T<4.45$ corresponding
to $0=m1<m2<m3$. As $T$ aproches $T_c=3.96$, $m2$ merges with $m1$.}
\label{freeenergy}
\end{figure}

In spite of the quantitative improvement of $T_c$, the order of the transition
is incorrectly given by the EDMF approximation. In Fig.\ref{magn3d} we
present result of solving EDMFT equations for the simplest $\pm 1$ spin
model in $d=3$. We plotted magnetization as a function of temperature. At
sufficiently low temperature the solution consists of three branches with
magnetizations $m_{3}>m_{2}>m_{1}=0$. These branches are extrema of
the free energy, which is shown schematically in Fig.\ref{freeenergy}.
$m_{3}$ and $m_{1}$ correspond to
local minima in the free energy, while $m_{2}$ corresponds to a local
maximum and is unphysical. The transition is clearly of the first order. The
order of the transition does not change up to $d=4$.

The inability of
EDMFT to predict the correct order of the transition is related to the
inability of a local theory to produce anomalous dimensions, and persists
in quantum problems when the dynamical critical exponent and the
dimensionality are such that they require the introduction of spatial
anomalous dimensions. 
Details are given in Appendix \ref{transition order} and \ref{largeNlimit}.

\bigskip

\section{Conclusion}

We have performed a semiclassical analysis of the EDMFT equations for a
simple Fermion Boson model. Comparison with earlier QMC treatments of the
same problem reveals that this method reproduces semiquantitatively all the
trends found in the previous study.\cite{Motome2000} It can be used
therefore in the study of more complicated systems, such as fermions
interacting with spin fluctuations.
We have also investigated this approach in the ordered phase revealing some
inadequacies of the approach which are closely related
to the existence of anomalous dimensions in finite dimensional systems.
Since this non trivial $k$ dependence which 
is characteristic of low dimensional
systems cannot be generated by a local theory, EDMFT\ produces spurious
results such as the existence of a first order phase transition in $d<4$ .
Since at zero temperature the dynamical critical exponent is such
that in two dimensions an ansatz without anomalous dimensions \ is
internally consistent, \cite{Si0012373} a continuation of the disordered
state,
beyond the first order phase transition, \ might be useful \ to study this
system.  In this spirit we pointed out that a continuation of 
the  EDMFT, at finite temperatures, suitably interpreted,  gives improved
estimates of the critical temperature compared to the simplest mean field
treatment or the Bloch Langer method.\cite{BlochLanger} It could be used to
obtain better estimates of ferromagnetic transition temperatures
where spatial fluctuations of the order parameter substantially decrease the
Curie temperature below the DMFT estimates. This is the case of bcc Iron,\cite
{iron} a problem which will require a more realistic investigation of
EDMFT.  Further investigation of the quantum problem will require zero
temperature methods which go beyond the semiclassical approximation.

\appendix

\section{Critical Temperature}
\label{ring diagrams}

The transition to the ordered state for a classical model
in BL method is signaled by the divergence in the zero
momentum term in the sum:
\begin{equation}
G_2=\sum\limits_q{\frac{J_q}{1-J_q M_2}}  \label{altedmft2}
\end{equation}
This equations is analogous to the selfconsistency equation
of the EDMFT, it is arises when summing ring diagrams. $M_2$
is a vertex and $J_{ij}$ is a line in the ring diagram.
$G_2(i) M_2$ is the sum of all ring diagrams which cover
the site $i$. $M_2$ and $G_2$ are related to $D$ in EDMFT:
\begin{equation}
D=M_{2}+M_{2}G_{2}M_{2}  \label{altedmft1}
\end{equation}

Below we are explicitly summing ring graphs on a Bethe
lattice to express $G_2$ through $M_2$.

We are introducing notations, $\tilde G_2(i)$ and $Q(i)$:
$\tilde G_2(i)$ equals $G_2(i)$ when the latter 
is computed on a lattice with all but one bonds cut out
from the site $i$. $Q(i) M_2$ includes 
those diagrams from $\tilde{G}_{2}(i)M_2$ which have only 
one vertex belonging to the site $i$.

The following relations can be established:

\begin{equation}
G_2=z\tilde G_2+z\tilde G_2 (z-1)M_2\tilde G_2 +z\tilde G_2 (z-1)M_2\tilde G
_2 (z-1)M_2\tilde G_2+...  \label{g2_series}
\end{equation}

\begin{equation}
\tilde G_2=Q+Q M_2 Q+Q M_2 Q M_2 Q+...  \label{tildeg2_series}
\end{equation}

\begin{eqnarray}
Q=J M_2 J+J M_2 (z-1)\tilde G_2 M_2 J+ J M_2 (z-1)\tilde G_2 M_2 (z-2)\tilde
G_2 M_2 J  \nonumber \\
+J M_2 (z-1)\tilde G_2 M_2 (z-2) \tilde G_2 M_2 (z-2)\tilde G_2 M_2 J+...
\qquad\qquad  \label{q_series}
\end{eqnarray}
where $z$ is the number of nearest neighbors, $J$ is a bond on the lattice.

Summing geometrical series, we obtain:

\begin{equation}
G_2={\frac{z\tilde G_2}{1-(z-2)M_2\tilde G_2}}  \label{g2}
\end{equation}

\begin{equation}
\tilde G_2={\frac{Q}{1-M_2 Q}}  \label{tildeg2}
\end{equation}

\begin{equation}
Q=J^2 M_2{\frac{1+M_2\tilde G_2}{1-(z-2)M_2\tilde G_2}}  \label{q}
\end{equation}

Solving these equations we get:

\begin{equation}
G_2={\frac{zQ}{1-zM_2Q}}  \label{g2_q}
\end{equation}

\begin{equation}
Q={\frac{1-\sqrt{1-(z-1)(2M_2 J)^2}}{(z-1)2M_2}}  \label{qsolved}
\end{equation}

Eq. (\ref{g2_q}) and Eq. (\ref{qsolved}) solve $G_2$ for $M_2$.

Curves $J_c$ vs $z$ for BL and EDMFT, together with MFT solution $J_c=1/z$
are presented in Fig.\ref{figtc} and compared to the exact solution:

\begin{equation}
J_c={\frac{1}{2}}\ln{\frac{z}{z-2}}  \label{exact}
\end{equation}

\section{Order of the phase transition}
\label{transition order}

Here we are proving that the EDMFT equations give \ a transition of the first
order for $d<4$ and of the second order in the higher dimensions. For
classical phonons EDMFT equations read:

\begin{eqnarray}
&m(r-J_{q=0})=-{\frac{\delta\Phi[m,D]}{\delta m}}  \nonumber \\
&D=\sum\limits_q[r+2{\frac{\delta\Phi[m,D]}{\delta D}}-J_q]^{-1}
\label{system}
\end{eqnarray}

It is easily seen that

\begin{equation}
\left.{\frac{\delta^2\Phi[m,D]}{\delta m^2}}\right|_{m=0}
=2{\frac{\delta\Phi[0,D]}{\delta D}}
\label{diff.relation}
\end{equation}

Solving Eq. (\ref{system}) for $r$ using the above relation for derivatives, 
up to the second order in $m$ we have:

\begin{equation}
-{\frac{m^{2}}{2}}D^{^{\prime \prime }}|_{m=0}=D|_{m=0}-\sum\limits_{q}
\left[
\left(2\frac{\delta ^{2}}{\delta D^2}
-\frac{1}{6}\frac{\delta^4}{\delta m^4}\right)
\left.\Phi[m,D]\right|_{m=0}m^{2}+J_{q=0}-J_{q}\right]^{-1}  
\label{uptom2}
\end{equation}
The coefficient in front of $m^2$ in the right hand side (rhs) is positive.
The lhs 
of Eq. (\ref{uptom2}) is $\propto m^{2}$; \ while the rhs has two
contributions, one $\propto m^{d-2}$ and the other $\propto \delta \beta $,
where $\delta \beta =\beta -\beta _{c}$. For $d<4$ the term $\propto m^{d-2}$
is dominant and $\delta \beta \propto -m^{d-2}<0$. A negative $\delta \beta $
implies the first order transition. For $d>4$ the term $m^{2}$ from lhs
becomes dominant and $\delta \beta \propto m^{2}>0$. This is the usual mean
field \ behavior resulting in a second order transition.

We showed that in a classical model the transition is of 
the first order below the upper critical dimension. The same
is true for a quantum transition as well. We show it in
the appendix \ref{largeNlimit} considering large $N$ limit. 

As discussed earlier in connection
with the order of the transition, this artifact of the EDMFT
results from the inability of a local theory to capture physics that
requires the introduction of anomalous dimensions. In spite of this
shortcoming,
when properly interpreted, EDMFT\ results in improved estimates of the
critical temperature relative to DMFT.

\section{Quantum phase transition}
\label{largeNlimit}

In this appendix we investigate the phase transition in the quantum
version of $\phi^4$ model. 
We compare EDMFT and a full lattice model using
large $N$ technique.
We will show that above the upper critical dimension $d>d_{uc}=4-z$
the exact critical exponents and the critical exponents obtained in
EDMFT coincide.
Below $d_{uc}$ the EDMFT and the lattice model exhibit different
critical exponents. In the EDMFT the transition is of the first order 
for ${1\over 2}d_{uc}<d<2$ and of the second order otherwise. 
The transition is of the second order in the lattice case. Moreover,
in EDMFT the exponents have universal value for $d<{1\over 2}d_{uc}$  
and a non universal value for ${1\over 2}d_{uc}<d<d_{uc}$.

The lattice model is described by the action:
\begin{equation}
\label{action}
S={1\over2}D^{-1}_0\phi^2+{U\over4}(\phi^2)^2
\end{equation}
where $D^{-1}_{0\omega,q}=r+|\omega|^{2\over z}+q^2$, 
$\phi^2=\sum_{a=1}^N\phi^2_a$, $U=u/N$, $r$ is a variable
parameter which drives the phase transition.
Corresponding EDMFT equations are:

\begin{equation}
mD_{0\omega,q=0}^{-1}+\frac{\delta\Phi[m,D]}{\delta m}=0
\label{edmft1}
\end{equation}

\begin{equation}
D=\sum_q[D_{0q}^{-1}+2\frac{\delta\Phi[m,D]}{\delta D}]^{-1}
\label{edmft2}
\end{equation}

The functional $\Phi[m,D]$ includes all two particle irreducible 
diagrams which are constructed from the magnetization $m$ (dot), 
the particle propagator $D$ (line) and the interaction term $U$ 
(four legged vertex). $\Phi$ satisfies the following equation:

\begin{equation}
\left.{\frac{\delta^2\Phi[m,D]}{\delta m^2}}\right|_{m=0}
=2{\frac{\delta\Phi[0,D]}{\delta D}}
\label{diffeq}
\end{equation}

Expanding $\Phi$ in small $m$ and using Eq. (\ref{diffeq}) we 
write EDMFT equations as:
  
\begin{equation}
D_{0\omega,q=0}^{-1}+2\frac{\delta\Phi[0,D]}{\delta D}+
{1\over6}\frac{\delta^4\Phi[m,D]}{\delta m^4}|_{m\to0}m^2=0
\label{edmft1exp}
\end{equation}

\begin{equation}
D_{\omega}=\int_0^{\Lambda} dq^d[|\omega|^{2\over z}+q^2+
\{2\Gamma-{1\over6}\frac{\delta^4\Phi[m,D]}{\delta m^4}|_{m\to0}\}
m^2]^{-1}
\label{edmft2exp}
\end{equation}  
where $\Gamma=\frac{\delta^2\Phi[0,D]}{\delta D^2}$.

Let $D_{0c}$, $D_c$, $r_c$ be values of $D_0$, $D$, $r$ in the transition
point.
Subtracting 
$D_{0c\omega,q=0}^{-1}~+~2\frac{\delta\Phi[0,D_c]}{\delta D}~=~0$ 
from Eq. (\ref{edmft1exp}) and keeping lowest order terms 
we have:

\begin{equation}
\delta r+2\Gamma\delta D
+{1\over6}\frac{\delta^4\Phi}{\delta m^4}m^2=0
\label{trans}
\end{equation}
where $\delta r=r-r_c$, $\delta D=D-D_c$. This equation provides a
relation between the variation of the driving term $r$ and the
order parameter $m$. We will show that for $d>d_{uc}$
the last term in the left hand side wins over the second term, 
the transition is mean field like. The second term becomes
important and determines the character of the transition
for $d<d_{uc}$.

\begin{figure}
\epsfxsize=6 truein
\epsfbox{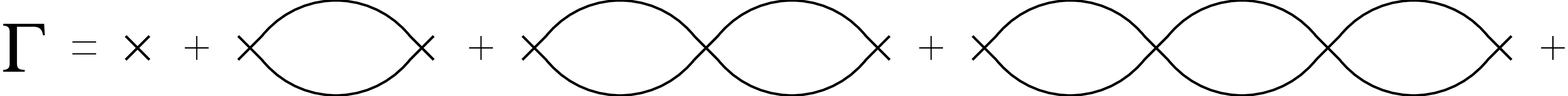}
\caption{$1/N$ expansion of $\Gamma$. All diagrams are of the
order $1/N$}
\label{Gamma.diag}
\end{figure}

We will consider the large $N$ limit up to the order of $1/N$. 
Diagrams which enter $\Gamma$ are chains of bubbles 
(see Fig.\ref{Gamma.diag}), which can be summed up as
geometrical series:

\begin{equation}
\Gamma={1\over N}\left(\frac{u}{2}+\frac{u}{1+\frac{u}{2}\chi}\right)
\label{Gamma}
\end{equation}
where $\chi_{\omega}\sim\int d\nu D_{\nu}
D_{\nu+\omega}$ in the quantum problem or $\chi=D^2$ in the
classical problem. 
The only term of the order $1/N$ which enters $\delta^4\Phi/\delta^4m$
is $6u/N$. All other terms are of order $O(1/N^3)$. 

Eq. (\ref{trans}) and Eq. (\ref{Gamma}) 
holds in case of a lattice as well, but $D$ now
depends on both momentum and frequency, and summations now
run over wave vectors as well.

The upper and lower critical dimensions are determined by
the convergence of integrals $Tr\delta D$ and $Tr D$ in
the ultraviolet and infrared limits respectively:
\begin{equation}
Tr\delta D\sim\int d\omega dq
\frac{q^{d-1}}{(|\omega|^{2\over z}+q^2)^2}
\label{int1}
\end{equation}

\begin{equation}
Tr D\sim\int d\omega dq
\frac{q^{d-1}}{(|\omega|^{2\over z}+q^2)}
\label{int2}
\end{equation}

These equations are the same for the mean field and lattice
models, they yield the upper critical dimension $d_{uc}=4-z$
and the lower critical dimension $d_{lc}=2-z=d_{uc}-2$.

We first consider EDMFT. In a crude way one can estimate:

\begin{eqnarray}
d>2,&\quad&D_{\omega}\sim\int d^dq
(|\omega|^{2\over z}+q^2)^{-1}
\sim(d-2)^{-1}(\Lambda_q^{(d-2)}-|\omega|^{d-2\over z})\\
d<2,&&D_{\omega}\sim
-(d-2)^{-1}|\omega|^{d-2\over z}
\end{eqnarray}
and
\begin{eqnarray}
d>2,&\quad&\chi_{\omega}\sim\int d\nu D_{\nu}D_{\nu+\omega}
\label{chi1}
\sim (d-2)^{-2}|\omega|^{{d-2\over z}+1}\\
d<2,&\quad&\chi_{\omega}
\sim -(d-2)^{-2}(2{d-2\over z}+1)^{-1}|\omega|^{2{d-2\over z}+1}
\label{chi2}
\end{eqnarray}
$\Lambda_q$ is a momentum cutoff.
We see from Eq. (\ref{chi2}) that for $d<d_{uc}/2$ the susceptibility
$\chi_{\omega}$ is divergent at low frequency, it
leads to a universal critical behavior for $d<d_{uc}/2$,
as follows from the self energy calculation below. 
The self energy in large $N$ limit is $\delta \Sigma
\sim 2\Gamma D$:
\begin{eqnarray}
{d_{uc}\over2}<d<d_{uc},&\quad&
\Sigma_{\omega}\sim {1\over N}(d-2)^{-1}u
|\omega|^{{d-2\over z}+1}\\
d_{lc}<d<{d_{uc}\over2},&\quad&
\Sigma_{\omega}\sim {1\over N}\int d\nu\chi^{-1}_{\nu}D_{\nu+\omega}
\sim {1\over N}(d-2)(2{d-2\over z}+1)|\omega|^{-{d-2\over z}}
\end{eqnarray}

In a similar way we can calculate a
contribution from $m$ to $\Gamma\delta D$ 

\begin{eqnarray}
{d_{uc}\over2}<d<d_{uc},&\quad&
\Gamma\delta D\sim {1\over N}(d-2)^{-1}u m^{d-2+z}\\
d_{lc}<d<{d_{uc}\over2},&\quad&
\Gamma\delta D
\sim {1\over N}(d-2)(2{d-2\over z}+1)m^{-d+2}
\end{eqnarray}

This result together with Eq. (\ref{trans}) suggests that the 
transition is the first order for ${1\over 2}d_{uc}<d<2$. 

Now we consider the lattice model.
\begin{equation}
\chi_{\omega,q}\sim\int d\nu d^dp
(|\nu+\omega|^{2\over z}+(p+q)^2)^{-1}(|\nu|^{2\over z}+p^2)^{-1}
\sim (d+z-4)^{-1}(|\omega|^{2\over z}+q^2)^{{d+z\over2}-2}
\label{chi_latt}
\end{equation}

\begin{equation}
\Sigma_{\omega,q}
\sim {1\over N}\int d\nu dp\chi^{-1}_{\nu,p}D_{\nu+\omega,p+q}
\sim{1\over N}(d+z-4)(|\omega|^{2\over z}\ln{|\omega|}+q^2\ln{q})
\label{Sigma_latt}
\end{equation}
In this case the frequency dependent part of the self energy
can be conveniently exponentiated to yield:
$D\sim [|\omega|^{2\over\tilde z}+q^{2-\eta}]^{-1}$ with 
$\tilde z=2-N^{-1}(d-d_{uc})c_1(d)$ and 
$\eta=-N^{-1}(d-d_{uc})c_2(d)$, where $c_1(d)$ and $c_2(d)$
are some smooth functions of $d$.

We also calculate a contribution from $m$ to $\Gamma\delta D
\sim N^{-1}(d-d_{uc})m^2\ln{m}$. 
It yields $\delta r\sim m^{1\over\beta}$
with $\beta={1\over2}+(d-d_{uc}){1\over N}c_3(d)$. The transition
is the second order in this case.

\section{Instability analysis}
\label{stability}

Let us consider a very general electron phonon Hamiltonian
which describes an electron phonon system with electron-electron
interaction (local or long range), electron-phonon interaction 
and phonon-phonon interaction (phonon unharmonicity). We can always
use a Hubbard Stratonovich decoupling on electron-electron 
interaction, so we assume that information about long range 
electron-electron interaction is stored in the phonon dispersion and
we will not write the long range interaction explicitly.  
We can introduce a source dependent action $S$ where the sources
are coupled to different fields. The free energy
$W=-\ln{\int e^S}$ is the generating functional for expectation 
values of those fields.

\begin{eqnarray}
\label{action_source}
S=\int & dx dx' &\nonumber\\
&c^{\dagger}_{\sigma}(x) & G_{0\sigma}^{-1}(x-x') c_{\sigma}(x')+
{1\over2}\phi(x)D_0^{-1}(x-x')\phi(x')
\nonumber\\
&& +\delta(x-x')\left(U n_{\uparrow}(x)n_{\downarrow}(x)+
V_4\phi^4(x)+\lambda\phi(x) c_{\sigma}^{\dagger}(x)c_{\sigma}(x)\right)
\nonumber\\
&&\quad\quad -J_{\sigma}(x,x')c_{\sigma}^{\dagger}(x)c_{\sigma}(x')
-{1\over2}\phi(x) K(x,x') \phi(x')
-\delta(x-x')L(x)\phi(x)
\end{eqnarray}
$x$ variable includes both space and time in the above formula
and repeated indices imply summation. Expectation values
of the fields coupled to the sources are given by:

\begin{equation}
\label{GDm}
G=\frac{\delta W}{\delta J},\,\,\,\,
K=2\frac{\delta W}{\delta K},\,\,\,\,
m=\frac{\delta W}{\delta L},\,\,\,\,
\end{equation}

Exact Green's functions correspond to the limit of zero sources.
To study phase transitions, like the transition when the phonon
field acquires non zero expectation value, one needs to have 
the free energy as a functional of correlation functions only.
Such functional can be derived as a Legendre transform of the
free energy: $\Gamma=W-JG-K/2 D-Lm$. The sources $J$,$K$ and $L$
have to be solved for $G$,$D$ and $m$. The functional $\Gamma$
is called a Baym Kadanoff functional and its stationarity yields
equations for zero source correlation functions. 
We present the functional without derivation:

\begin{eqnarray}
\label{BKfunct}
\Gamma_{BK}[G,D,m] & = & \nonumber\\
{\rm Tr}\log G & - & {\rm Tr}(G_0^{-1}-G^{-1})G
-{1\over2}{\rm Tr}\log D +{1\over2}{\rm Tr}D_0^{-1}D
+{1\over2}mD_0^{-1}m+\Phi[G,D,m]
\end{eqnarray}

$G_0$ and $D_0$ are free fields of the action, $\Phi$ functional is
the sum of all two particle irreducible graphs constructed from
the original bare interaction vertices, from vertices 
generated by shifting the phonon field by $m$ and from full 
correlation functions $G$ and $D$.

The charge ordering instability can be studied by looking
at the zero frequency-momentum phonon propagator behavior:
the propagator diverges in a charge density wave (CDW) 
transition. Alternatively
one can study the transition from the ordered side, by
observing the order parameter vanishing ($m$ in our case).
The two approaches should give consistent results. We will
first show that this is indeed the case in the exact theory,
then we explain how a similar approach can be applied in
the EDMF theory.

Let us introduce some compact notations we are going to use.
$o^a_{\alpha}$ is a field operator of $a$ kind at
$\alpha$ space-time point. $a=G$ specifies an electron
field operator and $a=D$ specifies a phonon field
operator.
$O^{ab}_{\alpha\beta,\gamma\delta}$ is a four point
function, which is a subset of all connected diagrams
in the perturbative expansion of 
$\langle o^{a\dagger}_{\alpha}o^a_{\beta} 
o^{b\dagger}_{\delta}o^b_{\gamma}\rangle$, rules
for selecting the subset of diagrams depends on
a particular operator. Multiplication of two operators
is defined by: $[O^{(1)}O^{(2)}]_{\alpha\beta,\gamma\delta}=
\sum_{\mu\nu}
O^{(1)}_{\alpha\beta,\mu\nu}O^{(2)}_{\mu\nu,\gamma\delta}$. 
We are introducing three four point operators:1) 
$\chi_0$ includes all graphs which enter skeleton graphs without
interaction vertices.
2) $\Sigma$ includes
all 1D irreducible diagrams; 3) $\Gamma$ includes all 2P
irreducible diagrams. In our case reducibility of 
$O_{\alpha\beta,\gamma\delta}$ is understood as disconnecting
$\alpha\beta$ part from $\gamma\delta$ part. ``1D irreducible''
means ``one particle irreducible with respect to cutting a
phonon line''. ``2P irreducible'' means ``two particle 
irreducible''. $\chi_0$ is trivially expressed in
terms of correlation functions: $\chi_0^{GD}=\chi_0^{DG}=0$,
$\chi_{0\alpha\beta,\gamma\delta}^{GG}=
G_{\alpha\gamma}G_{\beta\delta}$ and 
$\chi_{0\alpha\beta,\gamma\delta}^{DD}=
D_{\alpha\gamma}D_{\beta\delta}+D_{\alpha\delta}D_{\beta\gamma}$.

We can write out the following Dyson equations for 
the components of $\Sigma$ operator:
\begin{eqnarray}
\nonumber
\Sigma^{GG}&=&\chi_0^{GG}+\chi_0^{GG}\Gamma^{GG}\Sigma^{GG}
+\chi_0^{GG}\Gamma^{GD}\Sigma^{DG}\\
\nonumber
\Sigma^{GD}&=&\chi_0^{GG}\Gamma^{GG}\Sigma^{GD}
+\chi_0^{GG}\Gamma^{GD}\Sigma^{DD}\\
\nonumber
\Sigma^{DG}&=&\chi_0^{DD}\Gamma^{DD}\Sigma^{DG}
+\chi_0^{DD}\Gamma^{DG}\Sigma^{GG}\\
\Sigma^{DD}&=&\chi_0^{DD}+\chi_0^{DD}\Gamma^{DD}\Sigma^{DD}
+\chi_0^{DD}\Gamma^{DG}\Sigma^{GD}
\label{Sigmaeqs}
\end{eqnarray}

or we could simply write:
\begin{equation}
\label{SigmaDysonEq}
\Sigma=\chi_0+\chi_0\Gamma\Sigma
\end{equation}

Solving for $\Sigma$ we find:
$\Sigma=[\chi_0^{-1}-\Gamma]^{-1}=
-(\partial^2\Gamma_{BK})^{-1}$.
Second derivative $\partial^2\Gamma_{BK}$ 
is 2~x~2 matrix defined by:
\begin{equation}
\label{secder}
(\partial^2\Gamma_{BK})_{ab}=\frac{\partial^2\Gamma_{BK}}
{\partial C_a\partial C_b}
\end{equation}
where $C$ is a two component vector: $C_G=G, C_D=D$.

\begin{figure}
\epsfxsize=6.2 truein
\epsfbox{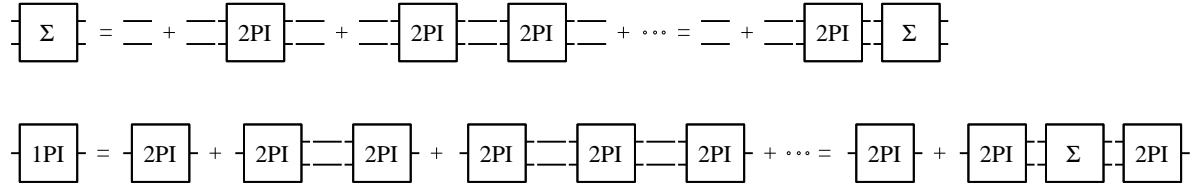}
\caption{Diagrammatic expansions for $\Sigma$ and $\Sigma_{ph}$}
\label{Sigmadiagrams}
\end{figure}

$\Sigma$ matrix is related to the phonon self energy $\Sigma_{ph}$
in a simple way, as can be seen from the diagrammatic series in 
Fig.\ref{Sigmadiagrams}.
$\Sigma$ comprises all four legged 1PI graphs, while $\Sigma_{ph}$
comprises all two legged 1PI graphs. 2PI four legged block 
is nothing but $\Gamma$.
Two horizontal lines represent a couple of correlation functions of the
same kind, $GG$ or $DD$, we assume that a summation runs over each couple of
horizontal lines, while 2PI four legged blocks are understood as 2 x 2
matrices. The first line is a diagrammatic analog of Eq. (\ref{SigmaDysonEq}).
The second line provides connection between $\Sigma$ and $\Sigma_{ph}$.
That can be written as:
\begin{equation}
\label{phSigma1}
\Sigma_{ph}=-\frac{\partial^2\Phi}{\partial m\partial m}+
\frac{\partial^2\Phi}{\partial m\partial C_a}
\Sigma^{ab}
\frac{\partial^2\Phi}{\partial C_b\partial m}
\end{equation}
or in a slightly different way:

\begin{equation}
\label{phSigma}
\Sigma_{ph}=-\frac{\partial^2\Phi}{\partial m\partial m}+
\frac{\partial^2\Gamma_{BK}}{\partial m\partial C}
\left(\frac{\partial^2\Gamma_{BK}}{\partial C\partial C'}\right)^{-1}
\frac{\partial^2\Gamma_{BK}}{\partial C'\partial m}
\end{equation}

The condition for the CDW instability at wave vector $q$ is: 
$D_{0q}^{-1}-\Sigma_{ph}=0$.

We will reproduce the above result studying CDW transition
from the ordered phase. $\Gamma_{BK}$ is the free energy, so
in the transition point 
\begin{equation}
\label{transition}
\frac{d^2\Gamma_{BK}}{dm dm}=0
\end{equation}
From the way $\Gamma_{BK}$ is constructed it follows 
$\partial \Gamma_{BK}/\partial C=0$ and $d\Gamma_{BK}/dm=
\partial\Gamma_{BK}/\partial m$. If we use 
$d(\partial \Gamma_{BK}/\partial C)/dm=0$
and Eq. (\ref{transition}) we find:
\begin{equation}
\label{transsolved}
\frac{\partial^2\Gamma_{BK}}{\partial m\partial m}-
\frac{\partial^2\Gamma_{BK}}{\partial m\partial C}
\left(\frac{\partial^2\Gamma_{BK}}{\partial C\partial C'}\right)^{-1}
\frac{\partial^2\Gamma_{BK}}{\partial C'\partial m}=0
\end{equation}
This equation is identical to Eq. (\ref{phSigma}) as should be
in exact theory. In EDMFT approach we take the local approximation
for the two particle irreducible graphs. All 2PI graphs in $\Gamma_{BK}$
are contained by $\Phi$. So the condition for $m$ vanishing is still 
given by Eq. (\ref{transsolved}) with $\Phi$ being local. Alternatively
we can use Eqs. (\ref{SigmaDysonEq}) where 
$\Gamma^{ab}=\frac{\partial^2\Phi}{\partial C_a\partial C_b}$ is local,
in which case these two methods are equivalent.
Let us consider the second method, when the transition is approached
from the disordered phase.

The local $\Gamma^{ab}$ can be computed using the 
impurity action of EDMF theory.
For simplicity we consider electron phonon interaction only, with
the coupling $\lambda=1$.
Equations similar to Eqs. (\ref{Sigmaeqs}) can be written for 
the susceptibility 
$\chi^{ab}=\langle o^{a\dagger} o^a o^{b\dagger} o^b\rangle$. 
In short notations it reads: 
\begin{equation}
\label{chiDysonEq}
\chi=\chi_0+\chi_0\tilde\Gamma\chi
\end{equation}
where $\tilde\Gamma$ is different from $\Gamma$ of Eq. (\ref{SigmaDysonEq}),
because now it includes 1D reducible diagrams. 
The relation between 
$\tilde\Gamma$ and $\Gamma$ is simple:
\begin{equation}
\label{GammaGamma}
\tilde\Gamma=\Gamma+\hat D_0
\end{equation}
where $\hat D_0$ is 2 x 2 matrix, $\hat D_0^{GG}=D_0$ and 
$\hat D_0^{DG}=\hat D_0^{GD}=\hat D_0^{DD}=0$.
Using Eq. (\ref{SigmaDysonEq}), Eq. (\ref{chiDysonEq}) 
and Eq. (\ref{GammaGamma})
we can express the self energy $\Sigma$ through the quantities
which are directly computed from the impurity action:
\begin{equation}
\label{Sigmasolved}
\Sigma=[[\chi_{imp}]^{-1}-[\chi_{0imp}]^{-1}
+\hat D_{0imp}+[\chi_0]^{-1}]^{-1}
\end{equation}
where $\chi_{imp}^{ab}=\langle o_a^{\dagger}o_ao_b^{\dagger}o_b\rangle_{imp}$,
$\chi_{0imp}^{ab}=\delta_{ab}C^2_a$ and $D_{0imp}$ is the Weiss field of the
impurity action:
\begin{eqnarray}
\label{impaction}
S_{imp}&=&\int d\tau d\tau'\\
&&c^{\dagger}_{\sigma}(\tau) 
G_{0imp,\sigma}^{-1}(\tau-\tau') c_{\sigma}(\tau')+
{1\over2}\phi(\tau)D_{0imp}^{-1}(\tau-\tau')\phi(\tau')+
\delta(\tau-\tau')\phi(\tau)c_{\sigma}^{\dagger}(\tau)c_{\sigma}(\tau)
\nonumber
\end{eqnarray}

The described method is exact in the limit $d\to\infty$. At finite $d$ it
yields a higher $T_c$ than a naive local approximation
$\Sigma_{ph}=\delta\Phi/\delta D$. Assuming 
$\Sigma_{ph}=\delta\Phi/\delta D$ would be equivalent
to taking $\Gamma_{BK}$ as being local in Eq. (\ref{phSigma}),
while the correct approach is to take a local approximation 
on the $\Phi$ functional only, not on the whole Baym Kadanoff functional.

\bigskip

ACKNOWLEDGMENT: This research was supported by the Division of Materials
Science of the National Science Foundation under
Grant DMR 89-15895-01.

\newpage
$\dagger$ Current address Institute of Materials Science, 
University of Tsukuba, Tsukuba, Ibaraki, Japan.

\end{document}